\documentclass[acmtog,nonacm]{acmart}

\AtBeginDocument{}
\setcopyright{none}
\authorsaddresses{}
\apptocmd{\maketitle}{\hypersetup{pdfcreator={LaTeX},pdfsubject={}}}{}{}

\usepackage{xcolor}
\usepackage{wrapfig}

\begin{document}

\title{Design for Machinability: Topology Optimization for Subtractive Machining Using Trimmed Neural Fields}

\author{Hongrui Chen}
\orcid{0009-0005-7465-8758}
\affiliation{%
  \institution{Carnegie Mellon University}
  \city{Pittsburgh}
  \state{Pennsylvania}
  \country{USA}
}
\email{hongruic@andrew.cmu.edu}

\author{Levent Burak Kara}
\orcid{0000-0002-2203-4020}
\authornote{Corresponding author.}
\affiliation{%
  \institution{Carnegie Mellon University}
  \city{Pittsburgh}
  \state{Pennsylvania}
  \country{USA}
}
\email{lkara@andrew.cmu.edu}

\newcommand{\Lc}{%
  \mathrel{\text{\scalebox{1.125}{\textcolor{cyan}{\textbf{\(\bigcirc\)}}}}}
  \kern-1.1em 
  \raisebox{0.15em}{\text{\scalebox{0.5}{\(\mathcal{L}_c\)}}}
}

\newcommand{\Lm}{%
  \mathrel{\text{\scalebox{1.125}{\textcolor{lime}{\textbf{\(\bigcirc\)}}}}}
  \kern-1.1em 
  \raisebox{0.15em}{\text{\scalebox{0.5}{\(\mathcal{L}_m\)}}}
}

\newcommand{\Lg}{%
  \mathrel{\text{\scalebox{1.125}{\textcolor{pink}{\textbf{\(\bigcirc\)}}}}}
  \kern-1.1em 
  \raisebox{0.15em}{\text{\scalebox{0.5}{\(\mathcal{L}_g\)}}}
}

\newcommand{\La}{%
  \mathrel{\text{\scalebox{1.125}{\textcolor{orange}{\textbf{\(\bigcirc\)}}}}}
  \kern-1.1em 
  \raisebox{0.15em}{\text{\scalebox{0.5}{\(\mathcal{L}_a\)}}}
}

\newcommand{\ch}[1]{{\color{black}{#1}}}
\newcommand{\chh}[1]{{\color{black}{#1}}}
\newcommand{\chhh}[1]{{\color{black}{#1}}}
\renewcommand{\shortauthors}{Chen and Kara}

\begin{abstract}
 
Topology optimization has found significant utility in additive manufacturing, but the resulting geometries are often difficult to realize using subtractive methods. The complex, curved surfaces of optimized designs typically require time-consuming point milling operations. In contrast, geometries composed of developable surfaces are well suited to flank milling, enabling substantial improvements in machining efficiency and surface quality. However, globally minimizing Gaussian curvature leads to overly simplistic designs. We address this challenge with a novel method that represents geometry using three implicit neural fields, optimized to produce machinable components. The resulting designs consist of regions of developable surfaces separated by seamlines where curvature is concentrated. Seamlines are efficiently computed as intersections of the implicit fields, while the geometry emerges through their product. By constraining surfaces to align with the cutter’s normal or tangential direction, our approach further accommodates practical constraints of 3-axis machining. We demonstrate concurrent optimization of topology and Gaussian curvature, enabling control over seamline feature scale. Compared to methods that enforce subtractive machinability without considering milling efficiency, our approach consistently reduces machining time while maintaining competitive structural performance.

\end{abstract}

\begin{teaserfigure}
  \includegraphics[width=\textwidth]{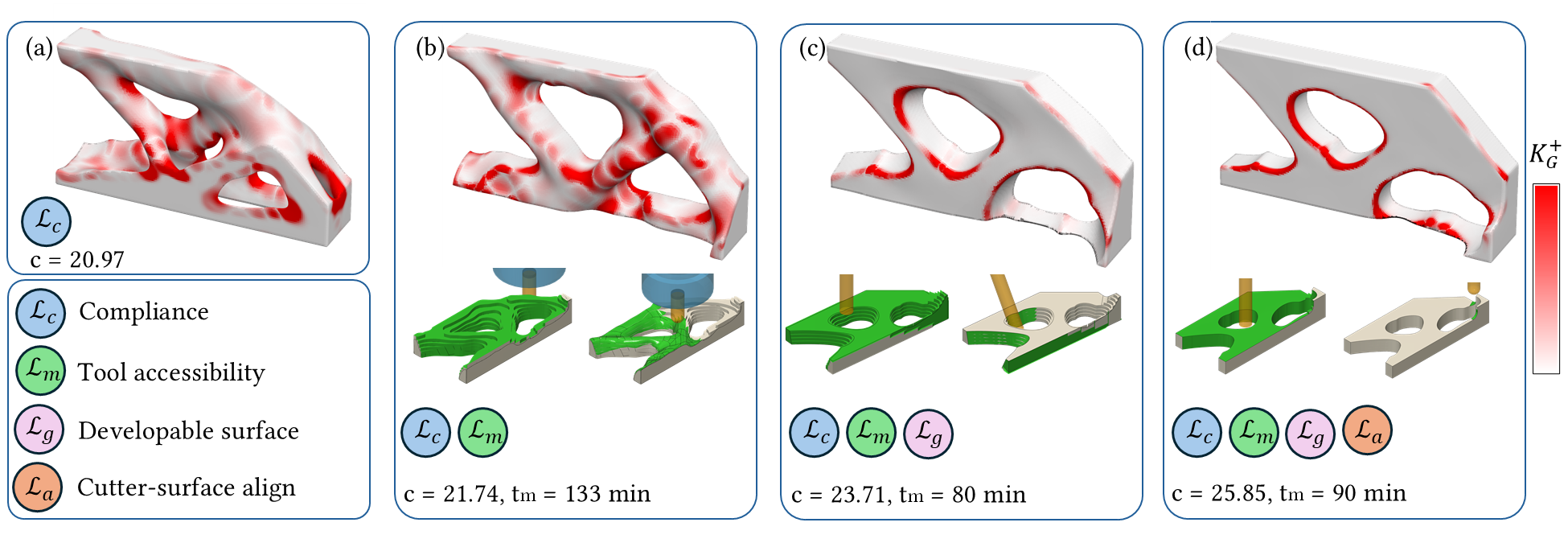}
  \caption{Our method for topology design for efficient machining using developable surfaces. (a) Topology optimization that minimizes compliance only. (b) Topology optimization with machining tool accessibility taken into account. Although machinable, such designs require time-consuming point milling to obtain an accurate realization of the surfaces. (c) Our method for generating developable surfaces is geared toward efficient flank milling. (d) A special case of our method tailored for 3-axis milling. Each design achieves a final volume fraction of 0.3, satisfying the constraint. The color map represents the absolute Gaussian curvature. Left-to-right: While machining becomes more efficient, geometries become increasingly more restricted (to accommodate machining goals), as reflected in the trade-off observed in the compliance.   }
  \Description{Four images illustrate designs from topology optimization and machining techniques: (a) A design minimizing compliance without machining considerations. (b) A design optimizing topology for tool accessibility, requiring point milling. (c) A design using developable surfaces optimized for efficient flank milling. (d) A special case of developable surface optimization tailored for 3-axis milling. Each design achieves a volume fraction of 0.3 and uses a color map to represent absolute Gaussian curvature. The progression from left to right shows increasingly constrained geometries for improved machining efficiency, accompanied by a trade-off in compliance.}
  \label{fig:teaser}
\end{teaserfigure}


\maketitle
\section{Introduction}

Topology optimization has emerged as a transformative tool for engineering design, enabling the creation of lightweight, high-performance structures by optimizing material distribution within a given design space. However, the intricate geometries generated by topology optimization often pose significant challenges for subtractive manufacturing processes such as milling. These geometries typically demand extended machining times and complex toolpaths to achieve acceptable surface finishes, making them poor choices for machinability. A promising alternative lies in leveraging developable surfaces, surfaces with zero Gaussian curvature, which are compatible with the highly efficient flank milling technique. Flank milling sweeps the cutter along the surface, which enables the creation of high-quality surface finishes while significantly reducing machining time \cite{he2021quasi}. Compared to conventional point milling, which only uses the tip of the cutting tool, flank milling offers a pathway to produce optimized designs that balance complexity and manufacturing efficiency. Despite this potential, no comprehensive method exists for topology design that accounts for the desire to produce developable surfaces. 

Current approaches that target topology optimization for subtractive manufacturing can be broadly categorized into those that rely on restrictive geometric constraints, post-processing techniques, or hybrid optimization methods. Methods that restrict designs to 2D extrusions or hierarchical 2.5D topologies, as seen in Autodesk Fusion 360’s topology optimization \cite{autodesk}, provide manufacturable designs but at the cost of reduced design freedom. Similarly, post-processing techniques attempt to minimally modify topology-optimized geometries into subtractive-compatible geometries but do not account for structural performance or load requirements during post-processing  \cite{stein2018developability}. Other methods aim to incorporate subtractive constraints directly into topology optimization \cite{langelaar2019topology,mirzendehdel2020topology} but often neglect machining efficiency, leading to designs with complex geometries that still require extensive machining time. In this work, we build upon these advances and describe a method that takes into account machining efficiency during topology optimization.  

A key to achieving our goal lies in the ability to produce designs with developable surfaces in a computationally tractable way. Directly minimizing the Gaussian curvature everywhere is ineffective, as the resulting shapes would consist of largely elementary building blocks. Moreover, computing accurate curvatures on finite element representations of topology-optimized shapes presents significant numerical challenges, particularly when generating the gradients of the objective functions. Additionally, generating designs whose surfaces align with subtractive manufacturing constraints, such as tool orientation and machinability, further complicates the problem.

We present a new approach based on implicit neural fields to address the challenges of incorporating developable surface constraints into the topology optimization process (Figure \ref{fig:teaser}). Our approach takes as input domain coordinates and outputs a density value at each coordinate, with the final design emerging as the intersection of three implicit fields. This representation enables the computation of analytical Gaussian curvatures.  The resulting geometry is characterized by regions of developable surfaces that join across seamlines where high curvature and directional changes are concentrated. By controlling the feature size of the implicit fields, our approach can regulate the spatial distribution of seamlines across the design domain, thereby facilitating the downstream conversion of implicit fields to solid geometries and the subsequent toolpath creation. Together with machinability constraints, our approach can create designs geared toward  5-axis flank milling to efficiently attain high geometric fidelity.  Additionally, our method can be tailored to  3-axis milling by incorporating mechanisms to control surface orientation, aligning surface normals with either tangential or normal cutting directions. By integrating these features, our approach ensures machining efficiency while maintaining design flexibility. We evaluate our approach across multiple case studies, comparing curvature, structural performance, and machining times. The results demonstrate that our approach consistently reduces machining times while incurring relatively small trade-offs in structural performance.

Our main contributions are: 
\begin{itemize}
    \item An approach to minimizing Gaussian curvature to generate developable surfaces suitable for efficient  5-axis machining.
    \item A novel tri-implicit field formulation to represent geometry, enabling precise seamline modeling and curvature concentration for machinable designs.
    \item An extension of curvature minimization techniques to align geometries with principal directions, enabling practical designs for 3-axis machining.
\end{itemize}

\section{Previous Work}
Our review focuses on topology optimization with implicit neural representations, topology optimization for subtractive manufacturing, and developable surfaces. 

\begin{figure*}[h]
  \centering
  \includegraphics[width=\linewidth]{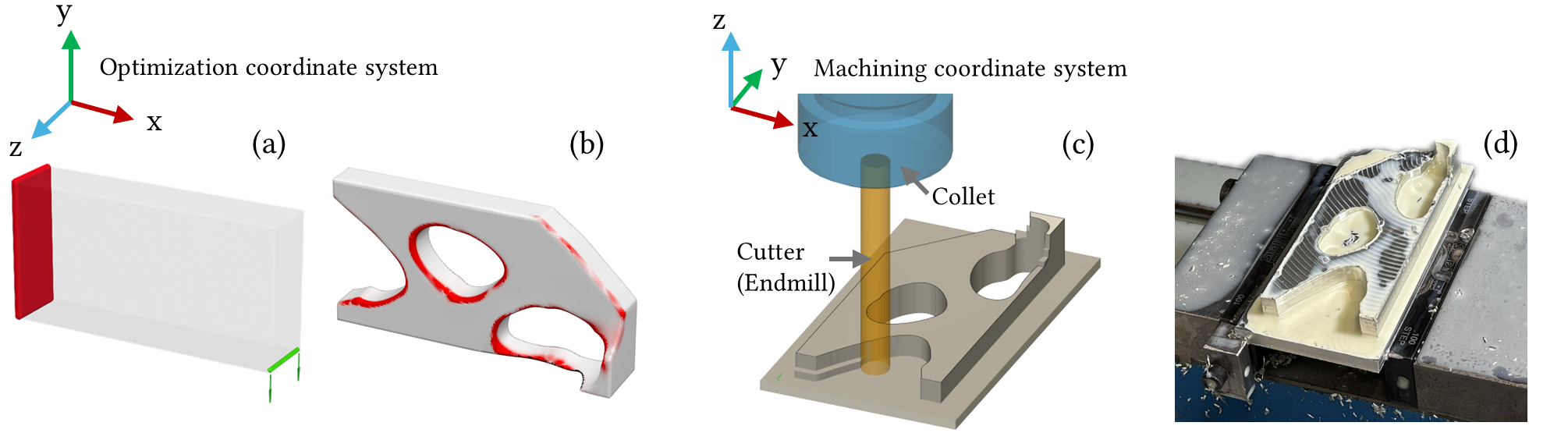}
  \caption{(a) Original design domain together with zero-displacement (red) and force (green) boundary conditions. (b) The designed part with an imposed 3-axis milling configuration. (c) The cutter relative to the workpiece in a 3-axis milling configuration. Note the coordinate axes relative to the cutter.  (d) Part being physically machined. }
    \Description{Four panels show: (a) the original design domain with fixed-displacement conditions in red and applied forces in green; (b) the designed part configured for 3-axis milling; (c) the cutter positioned relative to the workpiece, with coordinate axes indicating its orientation; and (d) the part being physically machined.}
  \label{fig:opt_cam_setup}
\end{figure*}

\textbf{Implicit neural representations}:
Implicit neural representations model geometry as an implicit function. Compared to mesh-based representations, neural fields can compactly encode shapes without the need for discretization while being infinitely differentiable. Prior work explored improving the representation capability of neural implicit fields \cite{atzmon2019controlling,deng2020nasa,gropp2020implicit,fan2024optimizing,petrov2023anise,marschner2023constructive}. Seamline modeling using implicit fields to represent vortex filaments \cite{ishida2022hidden}. Neural implicit representation also facilitates geometric deformations \cite{bozic2021neural,chen2021snarf}. Developments in implicit fields have also advanced high-quality rendering techniques \cite{sharp2022spelunking,takikawa2021neural,wang2023adaptive}, leveraging their ability to represent complex geometry compactly and with high precision. Additionally, applications have extended to font rendering, where three implicit fields are utilized for efficient and detailed representation \cite{chlumsky2018improved}. The differentiability of implicit fields enables accurate computation of normals and curvature, which is critical for downstream geometric processing tasks \cite{yang2021geometry,berzins2023neural}.


\textbf{Topology optimization with machine learning:} Topology optimization is a computational design methodology that iteratively distributes material within a given domain to meet specified performance objectives and constraints. Coupling finite element analysis with advanced optimization algorithms enables the creation of lightweight and structurally efficient solutions suitable for a broad range of engineering applications. Classical topology optimization includes SIMP \cite{bendsoe1989optimal,zhou1991coc} and the level-set \cite{allaire2002level} methods. Recent advances have enabled the generation of increasingly complex and intricate structures, with the resulting shapes being naturally compatible with additive manufacturing  \cite{liu2018narrow,zhu2017two,bickel2023design}. Recent approaches to integrating machine learning with topology optimization include data-driven topology optimization and direct topology optimization methods. In data-driven topology optimization, a neural network aims to learn from a database of already optimized designs. The neural network models include convolutional \cite{banga20183d,aage2015topology,nakamura2020deep,behzadi2021real,zheng2021generating}, generative adversarial  \cite{yu2019deep,nie2021topologygan} and diffusion-based approaches \cite{maze2022diffusion}. All these data-driven approaches focus on reducing the time required for optimal topology prediction. Although there have been improvements over the years, producing a widely applicable training set spanning many different boundary conditions is not straightforward, especially for 3D models. 

Another approach to neural topology optimization involves direct methods that do not use prior data but rather use a neural network to optimize the material distribution in the design domain  \cite{Chandrasekhar2021,Chandrasekhar2021Fourier,deng2020topology,deng2021parametric,zehnder2021ntopo,hoyer2019neural,han2024motion}. Such direct methods use spatial coordinates as input and produce an implicit field to represent the shape. By manipulating the spatial resolution of the input coordinates, length-scale control is achieved \cite{Chandrasekhar2021Fourier}.  Through automatic differentiation, the analytical density gradient can be accurately computed and integrated into optimization \cite{chen2023concurrent}. In this work, we build on these advances and use a neural implicit field for compact Hessian and  Gaussian curvature calculations.

\textbf{Subtractive manufacturing oriented topology optimization:} Subtractive machining is a widely used method in manufacturing. Starting from a block of material, it involves a cutting tool that shaves off material until the desired geometry is reached. To date, subtractive manufacturing-oriented topology optimization methods have primarily sought to ensure that the cutting tool (mill) can access all points of the designed part's surface, taking into account criteria such as the mill and collet interference, toolpath developability and other geometric constraints \cite{langelaar2019topology, vatanabe2016topology, liu20153d,morris2020subtractive,mirzendehdel2020topology,chenyh2016topology,liujk2017topology}.  However, these methods do not consider machining efficiency, where resulting shapes typically require time-consuming point milling to produce the optimized part.  Commercial software also demonstrated the stacking of multiple 2D extrusions to approximate designs that are amenable for 3-axis milling \cite{autodesk}. However, these extrusion-based methods further constrain the freedom of design. In this work, we retain the subtractive manufacturing constraints by Langelaar \cite{langelaar2019topology} to ensure accessibility of the tool head while incorporating a developable surface that can be efficiently machined with flank milling.

\textbf{Developable surfaces:} Developable surfaces have zero Gaussian curvature everywhere and can be flattened onto a plane without distortion, reflecting their roots in classical differential geometry \cite{yuan2023survey,hilbert2021geometry} They are generated using straight lines (rulings) \cite{wang2014multi,chu2008computer} and find utility in origami \cite{peraza2014origami,turner2016review,lee2018origami}, architecture \cite{nejur2016bringing}, and industrial design \cite{liu2011industrial,erdine2019curved,barda2023generative,kilian2008crease} due to their cost-effective production from sheet materials. If the original shape is developable, prior works on shape optimization \cite{schulz2017interactive, liu2020computer}, and beam-based \cite{morris2023bento, zhao2021bars} can also produce developable surfaces. Developability optimization on shapes aims to facilitate the creation of geometries that can be efficiently manufactured \cite{stein2018developability,sellan2020developability,rowe2024sharpening}. Our work is inspired by the work of Stein et al. \cite{stein2018developability} that demonstrates how a conventional topology-optimized shape can be post-processed to make it amenable to flank milling. In this work, we extend this insight, where we impose surface developability as a goal during the \textit{design} of the part rather than as a post-processing step. This allows our method to incorporate the knowledge of the boundary conditions and target mass criteria. To this end, we introduce a continuous neural representation of the geometry in the form of implicit fields, which facilitates differentiable geometry queries to drive topology optimization. 

To the best of our knowledge, our approach represents the first attempt at combining topology optimization with developable surface generation using implicit neural fields, with the ultimate goal of making subtractive machining more efficient. The next sections detail our approach and demonstrate our results. 

\begin{figure*}[h]
  \centering
  \includegraphics[width=\linewidth]{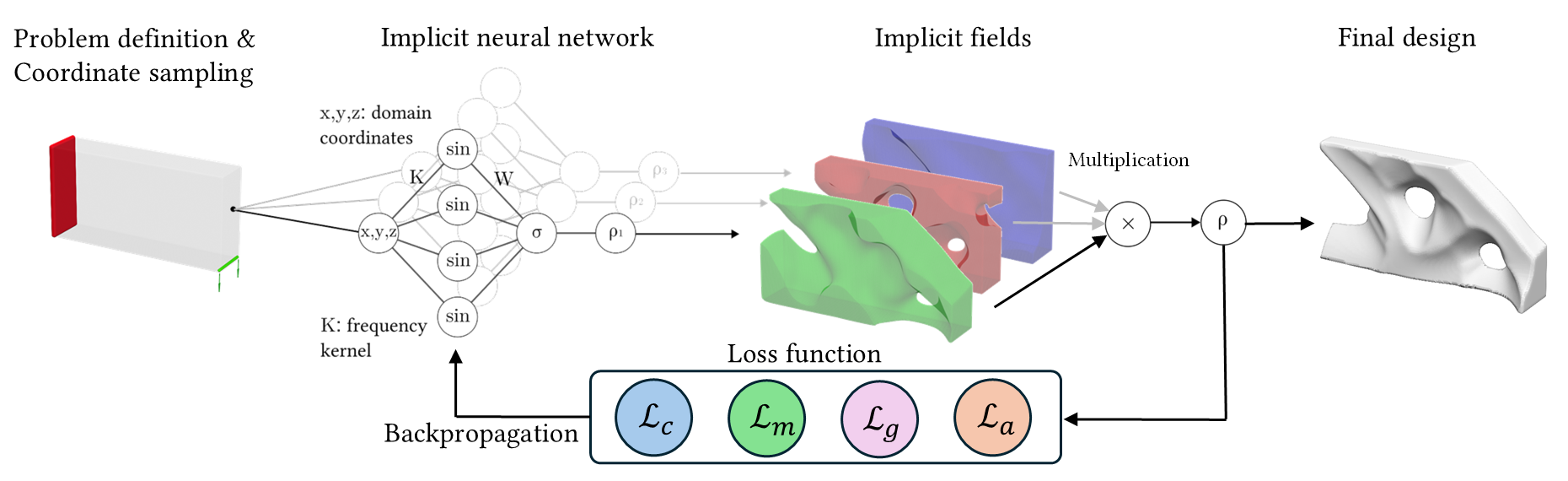}
  \caption{ The density field is represented as the product of three implicit fields. We sample coordinates across the design domain; the coordinates are then fed into three separate neural networks. The forward pass produces an output geometry. Various loss functions are calculated from this geometry and backpropagated to optimize the shape.    }
  \Description{Diagram showing a density field as the product of three implicit fields. Sampled coordinates are fed into three neural networks, producing an output geometry. Loss functions are computed and backpropagated to optimize the shape.}
  \label{fig:flowchart}
\end{figure*}

\section{Methodology}

Our approach aims to achieve concurrent topology optimization with developable surfaces to facilitate efficient subtractive machining. We limit our approach to structural compliance minimization under volume fraction constraints \cite{bendsoe1988generating,allaire2002level}. We build our approach based on a set of characteristics unique to subtractive machining, and the related objectives pertaining to its efficiency. We use an implicit neural representation to compute the Gaussian curvature, model seamlines through field intersections, and perform topology optimization. The optimization is performed by a neural network that handles gradient computations and backpropagation via automatic differentiation. This section details how these components integrate to optimize structural performance and improve manufacturability.

\subsection{Subtractive Machining}

A typical subtractive machining process consists of a roughing stage where the workpiece is cut close to the target geometry, followed by a finishing stage where surfaces and detailed features are machined to precision \cite{dong1993optimal,li1994optimal, lee1991cascam, kim2013rough}. For roughing, a single machining orientation (fixturing) is commonly used as it is both time and resource-efficient. In particular, most roughing operations are conducted using 3-axis machining due to the ease of programming, collision management, and faster execution times \cite{jousselin20195}. Consistent with the conventions in practice, we use $z-$ to denote the cutter's linear axis of motion (Figure \ref{fig:opt_cam_setup}) and assume the part will be fixtured \emph{only once} throughout the machining steps. Nonetheless, advanced methods that integrate roughing and finishing operations that utilize 5-axis joint roughing and finishing \cite{jousselin20195, dutta2024vector, mahdavi2020vdac} could be adopted for our Computer-Aided Manufacturing (CAM) simulations without loss of generality and without impacting our optimizer. We leave this option for future work. Additionally, in this work, we use Fusion 360 \cite{autodesk} as our CAM engine, which currently does not support 5-axis coordinated roughing. 

For finishing, we take into account 5-axis and 3-axis machining, which are supported by our CAM engine. Our proposed developable surface generation permits 5-axis flank milling, which we simulate using finishing operations. This enables the generation of developable surfaces to be cut using 5-axis milling, as shown in Figure  \ref{fig:teaser}(c). Similar to roughing, we assume a single fixturing of the part, hence excluding solutions that may be generated through multiple fixturing of the workpiece. Additionally, we do not consider the cutter's diameter on feature reproducibility, or the size of the cutter/collet on potential tool-part interference.

\subsection{Implicit neural representation}

Multiple studies have explored direct topology optimization with neural networks \cite{Chandrasekhar2021, Chandrasekhar2021Fourier}. The neural network replaces the finite element mesh as the geometry representation. During the optimization, the network is iteratively trained to produce a geometry that minimizes the prescribed loss function. In addition to representing geometry, the neural representation has the additional benefit of being an infinitely differentiable function. During each iteration, accurate gradients and  Hessians can be computed through automatic differentiation. Furthermore, seamlines can be modeled as the intersections of the surfaces implied by the implicit neural networks. 

\begin{figure}[h]
  \centering
  \includegraphics[width=\linewidth]{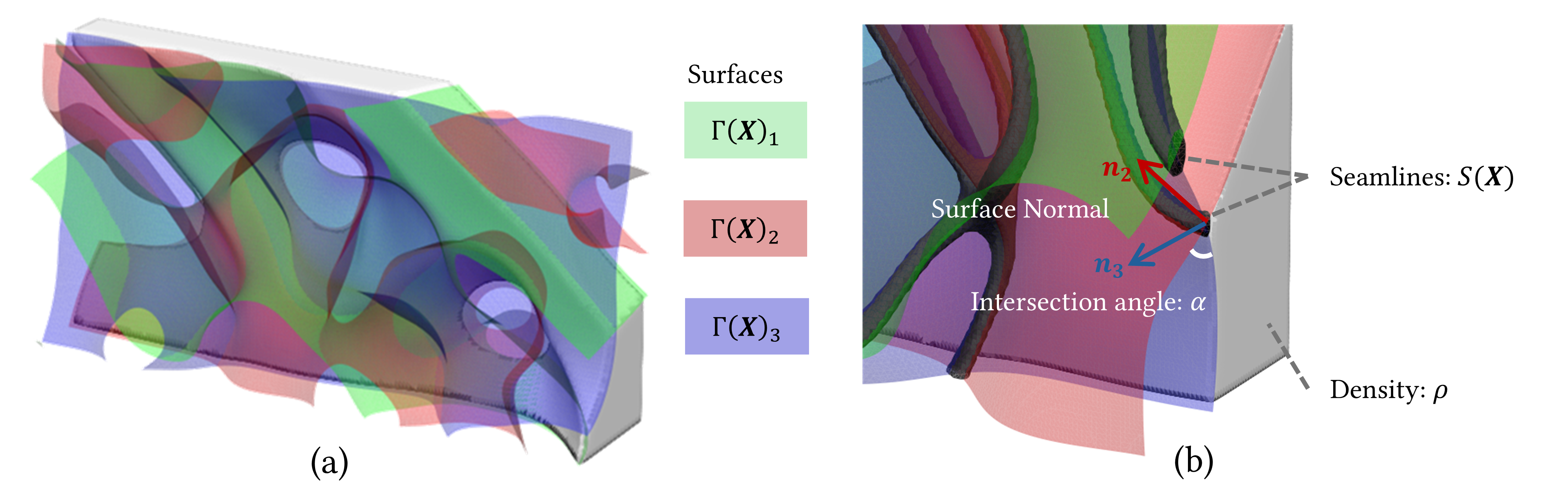}
  \caption{(a) Surfaces generated by the three networks. (b) Seamlines $S(\textbf{X})$ are formed at the intersections of surfaces. The seamlines are excluded from Gaussian curvature minimization. At a seamline, the two surfaces intersect at an angle $\alpha$. Large values of $\alpha$ indicate sharper seams, and thus fewer regions excluded from curvature minimization.}
  \Description{Boundary condition of the four test cases}
  \label{fig:field_intersection}
\end{figure}


\chh{
Figure \ref{fig:flowchart} illustrates the workflow. We employ a single-hidden-layer architecture that takes domain coordinates $\textbf{X} = [x, y, z]$, scaled to [-1, 1] along the longest dimension, and maps them to 512 frequency kernels. These are then linearly combined to produce a single output channel representing the density value $\rho$. This results in 2048 trainable parameters per field, comprising 512 $\times$ 3 = 1536 kernel parameters and 512 linear weights. The network capacity can be adjusted by varying the number of parameters.
}

\begin{equation}
\rho(\textbf{X}) = F(\textbf{X}) =\prod_{i=1}^n \sigma(\textbf{W}_i\sin(\textbf{K}_i\textbf{X} +  1)) 
\end{equation}

\noindent where:\\
$\rho$: Density field\\
$\textbf{X}$: Domain coordinate input, $\textbf{X}=(x,y,z)$ \\
$n$: Number of implicit fields, we use 3 \\
$\sigma$: Sigmoid activation function \\
$\textbf{K}$: Trainable frequency kernels \\
$\textbf{W}$: Trainable weights, initialized to 0\\

The network relies on the sinusoidal activation function, where similar implicit neural networks have been used for length scale control in topology optimization \cite{Chandrasekhar2021Fourier}. \chh{The three implicit fields play identical functional roles in the formulation and are not tied to fixed, predefined portions of the geometry. Each field defines a candidate implicit surface, and the final solid is obtained through the product of the three field responses. Accordingly, an individual field contributes by shaping the patch arrangement of the final piecewise developable geometry, rather than by representing a specific named component.}

\begin{wrapfigure}{r}{0.3\linewidth}
  \vspace{-10pt}
  \centering
  \includegraphics[width=\linewidth]{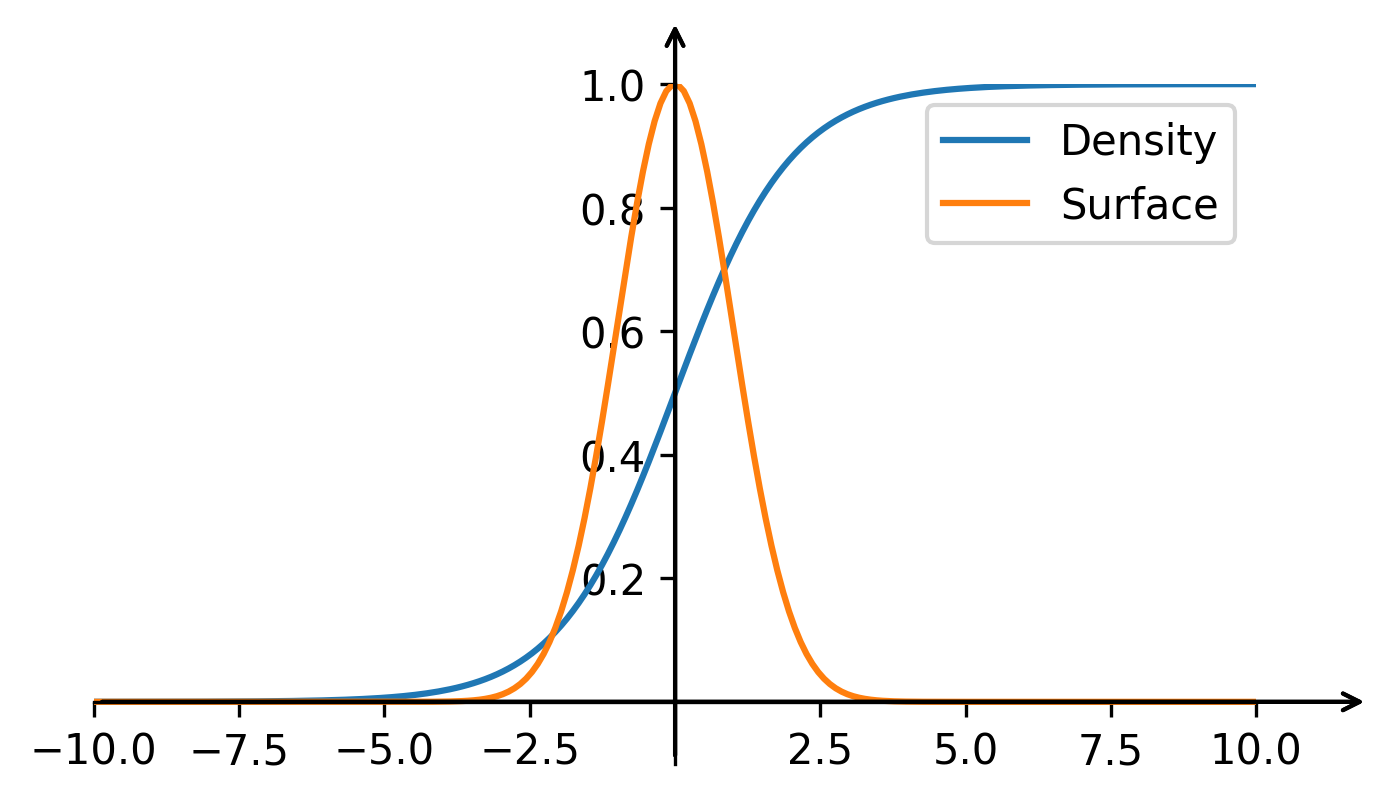}
  \Description{A plot compares a blue sigmoid density curve, which rises from zero to one, with an orange bell-shaped surface indicator centered at the transition point.}
    \vspace{-20pt}
\end{wrapfigure}
\noindent The Sigmoid activation function ensures $ 0 \leq \rho \leq 1$. \chh{With trainable weights initialized to zero, each network outputs a uniform density of 0.5 at initialization, resulting in $\rho(\textbf{X}) = 0.125$. This value is close to several of the target volume fractions considered in the paper.} In total, the neural network consists of one hidden layer. While deeper architectures that produce more complex maps can be utilized, this shallow architecture provides training efficiency.

To extract the surface of the geometry represented by each neural network, one can modify the activation function to obtain:

\begin{equation}
    \Gamma(\textbf{X})_i = e^{(-0.5\textbf{W}_i\sin(\textbf{K}_i\textbf{X} +  1))^2}
\end{equation}

Figure \ref{fig:field_intersection} illustrates $\rho(\textbf{X})$ vs. $\Gamma(\textbf{X})$. \ch{The surface is extracted at $\Gamma=1$.} With the surface $\Gamma(\textbf{X})$ of the $i^{th}$  network, we can compute the seamlines as the intersection between each surface pair based on the second elementary symmetric polynomial. We further filter the summation with a smooth Heaviside filter to unite the seamlines. The resulting seamlines $S(\textbf{X})$ are as follows:

\begin{equation}\label{eqn:seamlines}
S(\textbf{X}) = H\left(\sum_{1\le j<k\le n} \Gamma(\textbf{X})_j \Gamma(\textbf{X})_k \right)
\end{equation}

\noindent where we use the Heaviside function whose parameters we empirically determine: 

\begin{equation}
    H(x) = \frac{1}{1 + e^{-2 \cdot 10 \cdot (x - 0.2)}}
\end{equation}

Note that $0 \leq S(\textbf{X}) \leq 1, \forall ~\textbf{X}$. We use the term seamline with a slight abuse of meaning. Since all calculations are point-wise, it would be more appropriate to use `seampoint', but we prefer `seamline' for its colloquial familiarity.

\ch{The pairwise surface intersection can exhibit small nonzero values away from genuine intersections, causing false seamlines. To mitigate these, we shift the Heaviside function by 0.2 so only real overlaps activate. This threshold was chosen empirically for the balance between suppressing artifacts and preserving true seamlines.}

In this work, we use three implicit fields to generate trimmed surfaces formed by their intersections, which are then divided into distinct patches.
This means that the surface patches are $n$-colorable, where $n$ is the number of neural implicit fields used in the construction ($3$ for our examples). According to the Heawood Conjecture (a generalization of the Four Color Theorem to higher-genus surfaces \cite{ringel1968solution}), some maps on higher-genus surfaces are not $n$-colorable for {\em any} fixed $n$ and thus, if such an arrangement of surfaces is desired, more implicit fields will be required. 
In practice, we have not observed any issues -- our optimization controls patch placement and adjacency, which avoids worst-case topologies and enables a three-color scheme that balances clarity and computational efficiency.
Furthermore, it is relatively straightforward to increase the number of implicit fields to produce more complex surface patch layouts should the application require it.

\subsection{Topology optimization with developable surfaces}
Our overall topology design method considers four main objectives. First is topology optimization that performs compliance minimization under a user-prescribed volume fraction target \cite{bendsoe1988generating,Chandrasekhar2021,allaire2002level,liu2014efficient}. The second goal is to produce topologies that can be subtractively machined based on a projection method for accessibility calculations. Third is  Gaussian curvature minimization and the associated concentration of large curvatures along seamlines geared toward multi-axis flank milling. Finally, if specifically 3-axis milling is preferred (which remains very popular due to the ubiquitous availability of the machines in many shops), a fourth goal is to align the surfaces to the normal or tangent direction of the cutter. 

\textbf{Loss \#1 - $\mathcal{L}_{c}$: Compliance for topology optimization}. We adopt the advances in neural network-based topology optimization \cite{Chandrasekhar2021,Chandrasekhar_auto,Chandrasekhar2021MM,chen2023concurrent,chen2024conditioning}. The network takes as input space coordinates and outputs a material pseudo-density value associated with the SIMP method \cite{bendsoe1988generating}. Traditionally, SIMP-based compliance minimization can be formulated as:

\begin{equation}
\begin{aligned}
    &\text{minimize }_\rho && c(\rho^{(i)}) = \frac{1}{2} \mathbf{u}^T \mathbf{K}(\rho^{(i)}) \mathbf{u}, \\
    &\text{subject to } && \mathbf{K}(\rho^{(i)}) \mathbf{u} = \mathbf{f}, \\
    & && V(\rho^{(i)}) = \frac{1}{N}\sum_i \rho^{(i)} \leq V^*, \\
    & && 0 \leq \rho^{(i)} \leq 1, \quad \forall i.
\end{aligned}
\end{equation}

\noindent where $c$ is the compliance, $\rho^{(i)}$ is the set of design variables defined on the sampled coordinates. $\mathbf{u}$ is the displacement vector, $\mathbf{f}$ is the external force vector,  $\mathbf{K}$ is the stiffness matrix, $V$ is the resulting volume fraction, constrained by a prescribed upper bound $V^*$. 

In this work, we use a linear elastic, hexahedral finite element model defined on a uniform cartesian grid. Our solver receives each element’s density $\rho^{(i)}$ ($i$ indexes nodes of the Finite Element (FE) grid) as input and provides both the compliance and its corresponding sensitivity relative to the input coordinates. Following \cite{Chandrasekhar2021}, we compose a loss function that combines compliance minimization and the volume fraction constraint for  topology optimization: 

\begin{equation}
    \mathcal{L}_{c} = \frac{c(\rho^{(i)})}{c_0}+\omega_{c}(\frac{V(\rho^{(i)})}{V^*}-1)^2
\end{equation}

\noindent where $c$ is the current compliance (strain energy), $c_0$ is a normalization factor obtained from the initial unoptimized design domain. The work in  \cite{Chandrasekhar2021} has shown that this formulation produces results virtually identical to those obtained via SIMP-based compliance minimization algorithms such as the optimality criteria method \cite{bendsoe1989optimal}.

\textbf{Loss \#2 - $\mathcal{L}_{m}$: Tool accessibility for subtractive machining}. For a part to be subtractively machined, all regions of the part need to be accessible by the milling tool. The accessibility of a point within the design domain is evaluated by projections performed in the direction of machining. For a rectangular prism commonly used as the initial workpiece, the machining orientation is typically aligned with the coordinate axes. As described earlier, we simply adopt $-z$ as the cutter's central axis. With this, we formulate a smooth approximation to the projection via a cumulative summation \cite{langelaar2019topology}: 

\begin{equation} 
\begin{aligned}
    a^{(i)}_{[p,q,r]} &=
    \begin{cases}
        \rho^{(i)}_{[p,q,1]}, & r = 1, \\
        a^{(i)}_{[p,q,r-1]} + \rho^{(i)}_{[p,q,r]}, & r > 1.
    \end{cases} 
    \\
    & p = 1 \dots n_x, \quad q = 1 \dots n_y, \quad r = 1 \dots n_z.
\end{aligned}
\end{equation}

\noindent where $a^{(i)}$ captures the `accessibility' of  node $i$ located at $[p,q,r]$. $\rho$ represents the current density field defined on the same FE grid with the domain size of $(n_x, n_y,n_z)$. We use a  smoothed Heaviside function $H(.)$ to normalize the accessibility field and make it congruent with the density field and form a dissimilarity index $M$ that measures the dissimilarity between field $a$ and $\rho$: 

\begin{equation}
    M = \sum_i H(a^{(i)})(1-\rho^{(i)})
\end{equation}

Note that $0 \leq M \leq n_x n_y n_z$. During optimization, we employ the augmented Lagrangian method for subtractive loss to ensure the quality of results. The resulting subtractive machining loss is formulated as:

\begin{equation}
    \mathcal{L}_m = \lambda M+\chh{\frac{1}{2}}\gamma M^{2}
\end{equation}

\noindent where $\gamma$ is gradually increased during optimization and $\lambda$ is initialized to 1 and updated as:

\begin{equation}
    \lambda \leftarrow \lambda + \gamma M
\end{equation}


\textbf{Loss \#3 - $\mathcal{L}_{g}$: Developable surface}. 
From the neural network's output, we define the surface as the level set $F(\textbf{X}) =0.5$ for alignment with the adopted convention of $0 \leq \rho \leq 1 $. However, since the Gaussian curvature of the implicit surface only depends  on the gradient $\nabla F$, the Hessian $\mathcal{H}(F)$, and the adjoint of the Hessian $\mathcal{H}^*(F)$ \cite{goldman2005curvature}, the following identities still hold for the neural network's output:  

\begin{equation}
\nabla F 
= \left( \frac{\partial F}{\partial x}, \frac{\partial F}{\partial y}, \frac{\partial F}{\partial z} \right)
= (F_x,\,F_y,\,F_z),
\end{equation}

\begin{equation}
    \mathcal{H}(F)
= \begin{pmatrix}
\frac{\partial^2 F}{\partial x^2} & \frac{\partial^2 F}{\partial x\,\partial y} & \frac{\partial^2 F}{\partial x\,\partial z}\\[6pt]
\frac{\partial^2 F}{\partial y\,\partial x} & \frac{\partial^2 F}{\partial y^2} & \frac{\partial^2 F}{\partial y\,\partial z}\\[6pt]
\frac{\partial^2 F}{\partial z\,\partial x} & \frac{\partial^2 F}{\partial z\,\partial y} & \frac{\partial^2 F}{\partial z^2}
\end{pmatrix}
= \begin{pmatrix}
F_{xx} & F_{xy} & F_{xz}\\
F_{yx} & F_{yy} & F_{yz}\\
F_{zx} & F_{zy} & F_{zz}
\end{pmatrix}
\end{equation}

\begin{equation}
\begin{gathered}
\mathcal{H}^{*}(F)
\equiv \operatorname{adj}\!\left(\mathcal{H}(F)\right)
=
\begin{pmatrix}
C_{xx} & C_{xy} & C_{xz}\\
C_{xy} & C_{yy} & C_{yz}\\
C_{xz} & C_{yz} & C_{zz}
\end{pmatrix},
\\[4pt]
\begin{aligned}
C_{xx} &= F_{yy}F_{zz}-F_{yz}^{2},
& C_{xy} &= F_{xz}F_{yz}-F_{xy}F_{zz},\\
C_{xz} &= F_{xy}F_{yz}-F_{xz}F_{yy},
& C_{yy} &= F_{xx}F_{zz}-F_{xz}^{2},\\
C_{yz} &= F_{xy}F_{xz}-F_{xx}F_{yz},
& C_{zz} &= F_{xx}F_{yy}-F_{xy}^{2}.
\end{aligned}
\end{gathered}
\end{equation}

Goldman then derives the following form of the Gaussian curvature  \cite{goldman2005curvature}:  

\begin{equation}
K_G = \frac{\nabla F \,\ast\, \mathcal{H}^*(F) \,\ast\, \nabla F^{T}}{\lvert \nabla F \rvert^4}
\end{equation}

Ultimately, the goal is to minimize $|K_G|$ everywhere. For this, we seek a solid structure consisting of patches of developable surfaces joined along seamlines. These seamlines are where large curvatures are designed to be concentrated, enabling directional changes in the generated surfaces. Such seamlines have to be identified and sharpened within the optimization iterations so that they can be excluded from subsequent Gaussian curvature minimizations. To generate developable surfaces during optimization, we compute the Gaussian curvature and convert it into a loss function by minimizing the absolute value filtered by the seamlines $S(\textbf{X})$ (see Eqn. \ref{eqn:seamlines}). 

\begin{equation}
    \mathcal{L}_g = (1-S(\textbf{X}))|K_G|
\end{equation}

\emph{Ensuring seamline quality:} As described previously, a seamline $S(\textbf{X})$ is a scalar-valued function that characterizes the directional difference in the surface normals of two surfaces (Figure \ref{fig:field_intersection}). The local normal vector to a surface is given by:

\begin{equation}
\mathbf{n} = - \nabla F
\end{equation}

We quantify the local alignment of two surfaces via their local unit normals:

\begin{equation}\label{eqn:normalvectordotproduct}
\cos(\alpha) = \mathbf{\tilde{n}}_1 \cdot \mathbf{\tilde{n}}_2
\end{equation}

\noindent where $\alpha$ is the angle between the two surfaces, $\mathbf{\tilde{n}}_1$ and $\mathbf{\tilde{n}}_2$ are the unit normals. 

Ideally, surfaces obtained from our parallel neural implicit fields would intersect at sharp angles, leaving large swaths of surface points to be available for $|K_G|$ minimization. As such, we seek to form sparse but strong seamlines that are characterized by the concentration of large directional changes. For two surfaces that have an intersection angle $\alpha$ greater than the critical intersection angle, the following inequality must hold true:

\begin{equation} \label{eqn:anglealphacriterion}
    |\cos(\alpha)| = |\mathbf{\tilde{n}}_1 \cdot \mathbf{\tilde{n}}_2|\leq \cos(\alpha_{cr})
\end{equation}

\noindent where  $0 \leq \alpha_{cr} \leq \frac{\pi}{2}$ is a tunable angle that enforces the sharpness of the seam. Large values of $\alpha_{cr}$ help eliminate the formation of overlapping (aligned) surfaces, thereby restricting seam formation to sparse and narrow portions of the identified surfaces. The absolute value $|\cos(\alpha)|$ accounts for intersection angles greater than $\frac{\pi}{2}$.


To produce sharp seams between two surfaces $\Gamma_i$ and $\Gamma_j$ in a differentiable way, we formulate a seam quality loss function by modifying Eqn. \ref{eqn:anglealphacriterion}:

\begin{equation}
    L_{s,ij} = |H(|\mathbf{\tilde{n}}_i \cdot \mathbf{\tilde{n}}_j|-\cos(\alpha_{cr}))\cdot\mathbf{n}_i\cdot\mathbf{n}_j|
\end{equation}

\noindent where H(.) is the Heaviside function. The unnormalized surface normals allow the loss function to amplify the contributions of $\textbf{X}$ that form the true surfaces (those that give rise to $\Gamma(\textbf{X})$) over those that reside off the identified surfaces.  For three implicit fields with normals $\textbf{n}_1$, $\textbf{n}_2$, $\textbf{n}_3$, the combined seam quality loss function becomes: 

\begin{equation}
\mathcal{L}_s(\textbf{X}) = L_{s,12}(\textbf{X}) + L_{s,13}(\textbf{X}) + L_{s,23}(\textbf{X})
\end{equation}

\textbf{Loss \#4 - $\mathcal{L}_{a}$: Cutter-surface alignment}. This is an optional loss in case one wants to further restrict the optimization for 3-axis machining. For 3-axis machining, the cutter can remove local material most efficiently when the cutter's axis is normal (end-milling) or tangent (flank-milling) to the surface. We encode this in the form of a normal alignment $A_n$, and tangent alignment $A_t$  loss. If we denote the cutter's rotational axis as $\mathbf{r}$ (which remains fixed relative to the world coordinates in 3-axis machining), we have:

\begin{equation}
\begin{split}
    A_n &= |H(|\mathbf{\tilde{n}} \cdot \mathbf{r}|-\cos(\beta_{cr}))\cdot\mathbf{n}\cdot\mathbf{r}|\\
    A_t &= |H(|\mathbf{\tilde{n}} \times \mathbf{r}|-\cos(\beta_{cr}))\cdot(\mathbf{n}\times\mathbf{r})|
\end{split}
\end{equation}

\noindent where $\mathbf{n}$ is the surface normal of the current density field, $H(.)$ is the Heaviside function, $\beta_{cr}$ (we gradually increment to 90$^\circ$) is a user-defined angle threshold above which misalignments are penalized.

For efficient 3-axis machining, the cutter  needs to align either with the normal or  the tangent direction of the surface, which  can be compactly written as an overall cutter-surface alignment loss:

\begin{equation}
    \mathcal{L}_a = A_n A_t
\end{equation}

It is interesting to note that this loss is complementary to $\mathcal{L}_g$. An open surface that at each point is either normal or tangent to a fixed direction $r$  necessarily has $K_G=0$ and is therefore developable. In the case of normal alignment where $A_n$ would be minimized, the surface normal remains constant, which forces the surface to be planar. In the case of tangent alignment where $A_t$ would be minimized, the surface is curved at most up to straight lines, all parallel to $r$, forming a cylindrical surface. Despite this strict alignment already implying developability, for 3-axis machining, we retain the original $\mathcal{L}_g$ to further assist with surface developability. 

\textbf{Combined loss}. The topology optimization $\mathcal{L}_c$, subtractive machining $\mathcal{L}_m$, Gaussian curvature with sharp seams {$\mathcal{L}_g$, $\mathcal{L}_s$}, and alignment loss $\mathcal{L}_a$ are then combined into a single loss function. 
\begin{equation}
    \mathcal{L} = \mathcal{L}_c + \mathcal{L}_m + \omega_g\mathcal{L}_g + \mathcal{L}_s + \mathcal{L}_a
\end{equation}

\begin{wrapfigure}{r}{0.4\linewidth}
  \vspace{-10pt}
  \centering
  \includegraphics[width=\linewidth]{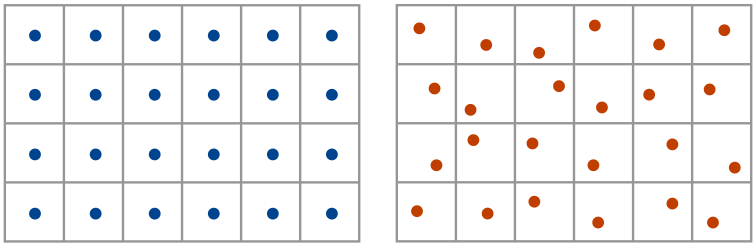}
  \Description{Two rectangular grids compare regularly aligned blue samples at cell centers with orange samples jittered to varying positions within the grid cells.}
    \vspace{-20pt}
\end{wrapfigure}
During optimization, we gradually increase $\omega_g$ to minimize $|K_G|$. $\mathcal{L}_a$ is only enabled if optimizing for 3-axis machining. We gradually increase $\alpha_{cr}$ and $\beta_{cr}$.  \chh{For $\mathcal{L}_c$ and $\mathcal{L}_m$, we sample at FE-grid-aligned coordinates (blue points located at the centers of FE cells), as compliance and accessibility are defined on the FE discretization and projection grid. For $\mathcal{L}_g$, $\mathcal{L}_s$, and $\mathcal{L}_a$, we employ quasi-Monte Carlo sampling implemented as jittered FE-grid samples (orange points distributed within the FE grid), with new random samples drawn at each epoch, since these geometric losses are not tied to the FE discretization. In all cases, the number of sampled coordinates per epoch is set equal to the number of FE grid nodes.}

Table \ref{tab:hyper_update} describes the configurations of various hyperparameters used in our study. 


\begin{table}[h!]
\centering

\begin{tabular}{l|c|c|c}
Hyperparameter & Starting epoch & $\Delta$ & Final value \\
\hline
$\omega_c$ & 0 & 0.5 & 100 \\
$\gamma$ & 20 & 0.033 & \chh{2} \\
$\alpha_{cr}$ [$^\circ$] & 60 & 0.267 & 80 \\
$\beta_{cr}$[$^\circ$] & 20 & 0.3 & 90 \\
$\omega_g$ & 40 & 0.033 & 1.0 \\
\end{tabular}
\caption{Our hyperparameter settings. \chh{All parameters start at 0}. Increments are uniform with a step size of $\Delta$ per epoch.}
\label{tab:hyper_update}
\end{table}

\section{Results and Discussion}

This section details several 3D optimization studies comparing the following approaches: (A) topology optimization aimed at minimizing compliance, (B) topology optimization incorporating constraints for machining tool accessibility, (C) our proposed method for generating developable surfaces through topology optimization, and (D) a specific implementation of our method optimized for 3-axis milling. We evaluate the structural performance and machining time estimates derived from computer-aided manufacturing (CAM) simulations. Furthermore, we compare CAM-simulated machining times with real, physical machining times to validate the findings. 

All experiments were conducted on a PC equipped with an Intel i7-12700K processor, 32 GB of RAM, and an NVIDIA RTX3080 GPU. \chh{We use the Adam optimizer \cite{kingma2014adam} in all experiments, with a learning rate of 0.002, no weight decay, and no gradient clipping. All experiments are run for 400 epochs.} The supplementary materials include the code implemented in TensorFlow \cite{Abadi2016} necessary to replicate these experiments.

\subsection{Test Cases, Geometric and Structural Performance Comparisons}

\begin{table*}[h!]
\centering

\begin{tabular}{l|c|c|ccc|ccc|ccc}
\textbf{Problem Type} & \textbf{FE Size} & \(V^*\) & \multicolumn{3}{c|}{\textbf{Compliance only $\Lc$ }} & \multicolumn{3}{c|}{\textbf{Subtractive accessibility $\Lc$ $\Lm$}} & \multicolumn{3}{c}{\textbf{Developable surface $\Lc$ $\Lm$ $\Lg$}} \\
 &  &  & \(V\) & \(c\) & \textit{runtime(s)} & \(V\) & \(c\) & \textit{runtime(s)} & \(V\) & \(c\) & \textit{runtime(s)} \\
\hline
\textbf{MBB} & $40\times20\times8$ & 0.3 & 0.301 & 20.97 & 650 & 0.301 & 21.74 & 608 & 0.301 & 25.81 & 608 \\
\textbf{Mount} & $40\times20\times8$ & 0.2 & 0.200 & 0.84 & 458 & 0.201 & 1.1 & 453 & 0.202 & 1.64 & 476 \\
\textbf{Pedal} & $50\times20\times8$ & 0.2 & 0.200 & 31.82 & 651 & 0.200 & 32.65 & 603 & 0.201 & 38.65 & 626 \\
\textbf{Link} & $40\times20\times8$ & 0.25 & 0.251 & 11.99 & 668 & 0.252 & 15.63 & 613 & 0.252 & 23.09 &  645\\
\end{tabular}
\caption{Analytics from the four test cases. We observe that compliance saw a slight increase with the additional loss functions. All runs reached within 1\%  of the target volume fraction. The runtime slightly varies, but we see no significant worsening in runtime with the additional loss functions. }
\label{tab:bc_compare}
\end{table*}

\begin{figure}[h]
  \centering
  \includegraphics[width=\linewidth]{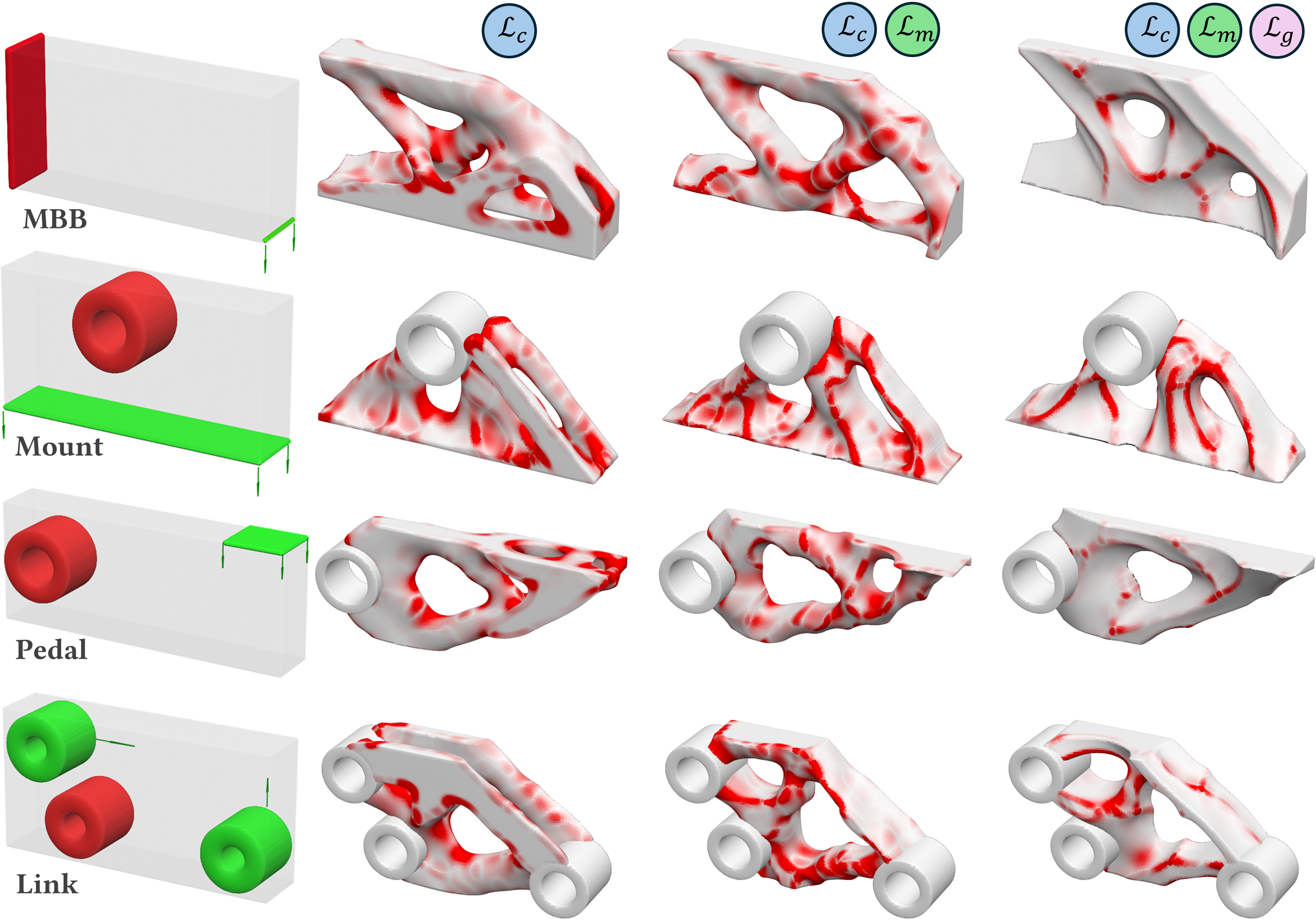}
  \caption{Rows represent different example problems. The first column illustrates zero-displacement boundary conditions (red) and external forces (green). The subsequent columns display the optimization results for various objective functions, with the final column showcasing designs incorporating the developable surface objective. Regions of high absolute Gaussian curvature $K_G$ are highlighted in red. Topology optimization with tool accessibility constraints (third column) yields complex surfaces characterized by extensive regions of high curvature. In contrast, our method (final column) generates sharper seamlines and larger areas of uniformly low Gaussian curvature, demonstrating improved surface regularity and manufacturability.}
\Description{Grid of optimization results for four test problems. The first column shows the boundary conditions, and the remaining columns compare compliance-only, subtractive-accessibility, and developable-surface optimization results.}
  \label{fig:bc_compare}
\end{figure}

Four test cases were utilized to assess the structural and manufacturing efficiency of the proposed method: (1) the MBB beam, a standard benchmark for topology optimization algorithms \cite{liu2014efficient}, (2) a bearing mount, (3) a pedal, and (4) a linkage structure. Figure \ref{fig:bc_compare} illustrates the boundary conditions for each case. When comparing compliance-only topology optimization to topology optimization with machining tool accessibility, the latter eliminates overhang regions by incorporating tool accessibility constraints. Similarly, the incorporation of developable surfaces further eliminates overhanging geometries. An analysis of surface curvature reveals that the developable surface optimization concentrates curvature along seamlines, leaving other areas with negligible Gaussian curvature.

\textbf{Convergence and performance}. Compliance was evaluated across the four test problems (Table \ref{tab:bc_compare}). Results indicate a progressive increase in compliance as more constraints are imposed, from compliance-only optimization to subtractive tool accessibility, and subsequently to developable surfaces. This is consistent with the expected trade-off between structural performance and manufacturability, with additional constraints negatively impacting compliance.

\begin{figure}[h]
  \centering
  \includegraphics[width=\linewidth]{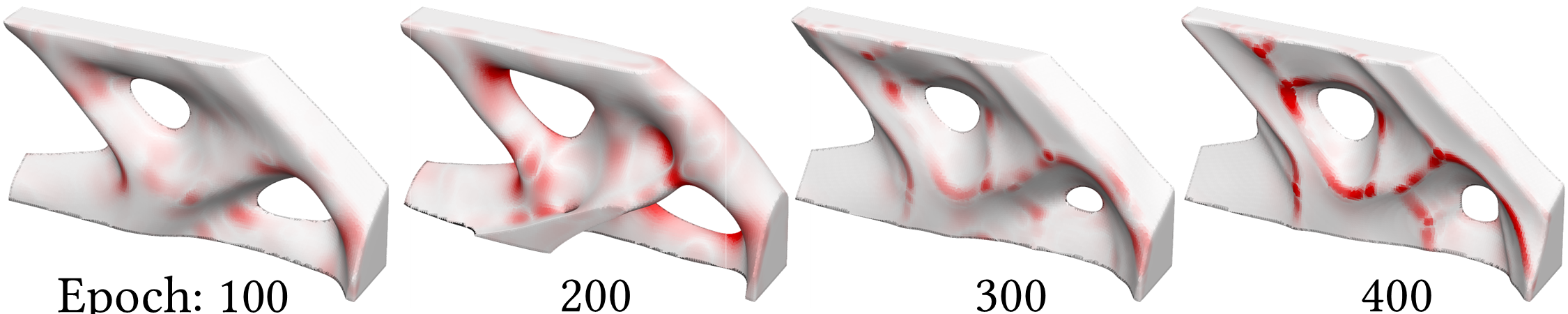}
  \caption{Convergence history for topology optimization incorporating developable surface objectives. The results demonstrate a progressive concentration of high curvature regions along the seamlines.}
  \Description{Convergence history for topology optimization incorporating developable surface objectives.}
  \label{fig:ds_converge}
\end{figure}

\begin{figure}[h]
  \centering
  \includegraphics[width=\linewidth]{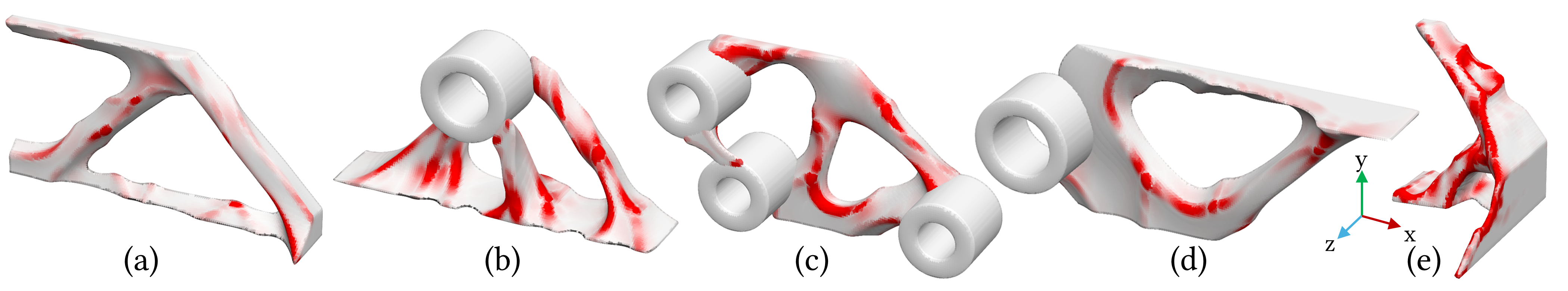}
  \caption{\ch{(a-d) Four test cases, and (e) an elongated z-axis example of the MBB. All problems enforce 0.1 volume fraction. With the lower volume fraction, the results still exhibit prominent seamlines.  }}
  \Description{Feature size control achieved through uniform scaling}
  \label{fig:01vf}
\end{figure}

\begin{figure*}[h]
  \centering
  \includegraphics[width=0.8\linewidth]{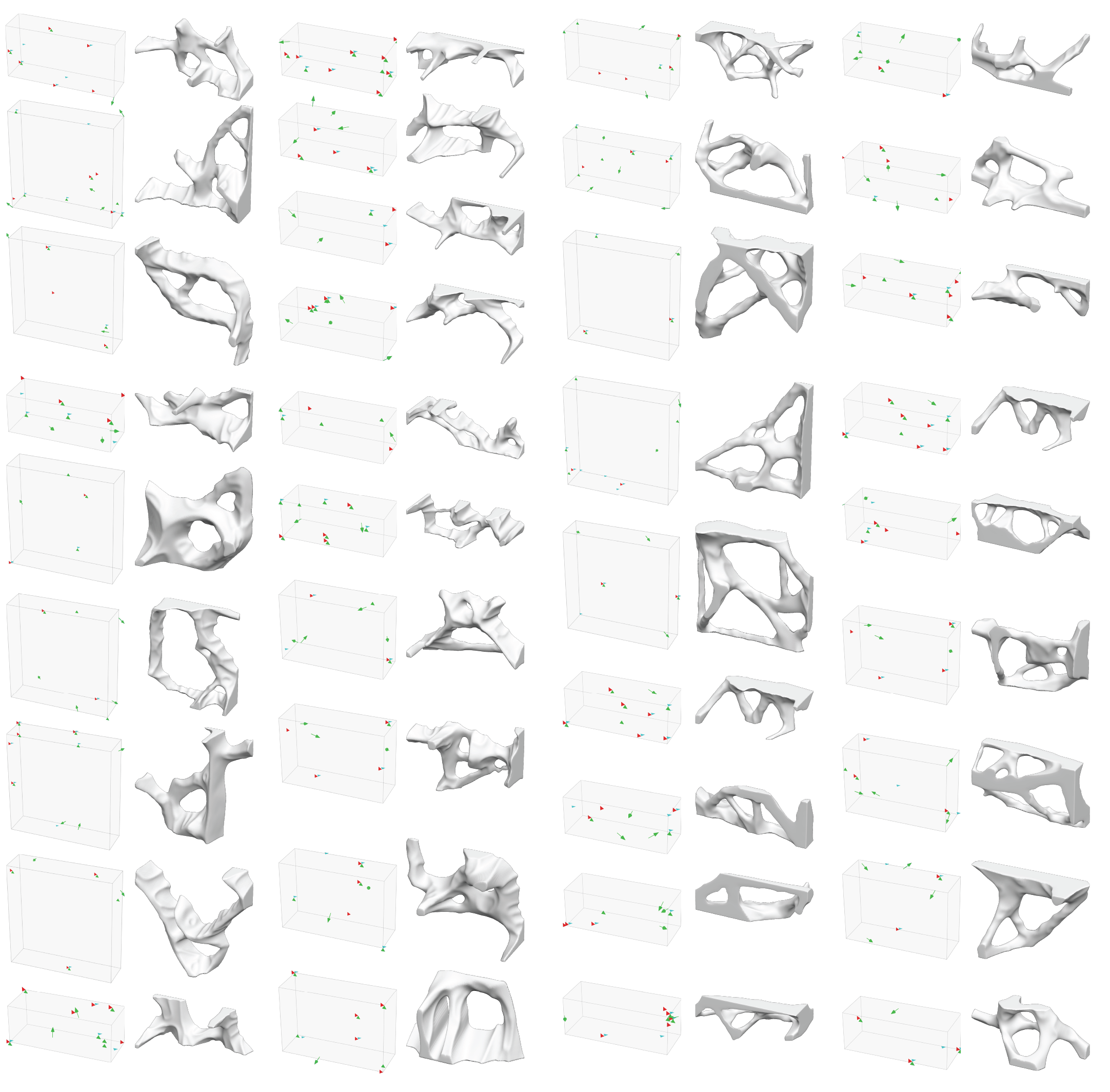}
  \caption{\chh{Our approach  is evaluated on varying design domain aspect ratios (gray boxes), randomized fixed displacement boundary conditions (red markers), and load cases (green markers), all with a low volume fraction of 0.1. Best viewed with digital zoom. For the left two columns, we use a single manufacturing orientation (z-), while the right two columns include both z+ and z- orientations.} \chhh{Relative to the compliance-only baseline, subtractive-only optimization and developable-surface optimization increase compliance by 45.82\% and 147.3\% on average, respectively, across the randomized-boundary-condition cases. }}
  \Description{Variation of z-axis coordinate scaling}
  \label{fig:random_bc}
\end{figure*}

\begin{figure}[h]
  \centering
  \includegraphics[width=0.9\linewidth]{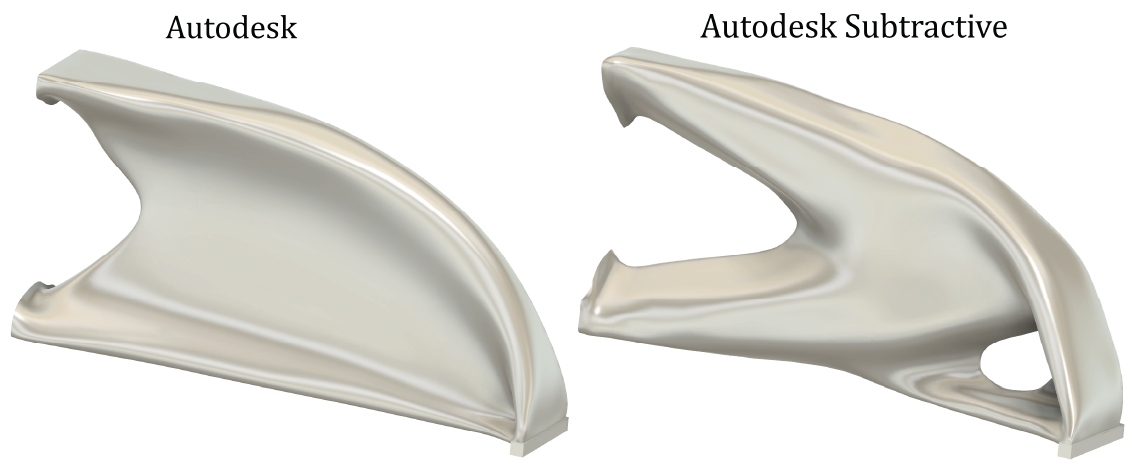}
  \caption{\chh{We compare against Autodesk Fusion 360’s generative design for a cantilever beam with identical boundary conditions and a target volume fraction of 0.3. The left shows unconstrained topology optimization in Autodesk Fusion 360, while the right shows generative design with a subtractive cutting constraint under a z- orientation in Autodesk Fusion 360. For this cantilever example, we do not observe significantly more complex features.}}
  \Description{Autodesk benchmark}
  \label{fig:f360}
\end{figure}

\begin{figure}[h]
  \centering
  \includegraphics[width=\linewidth]{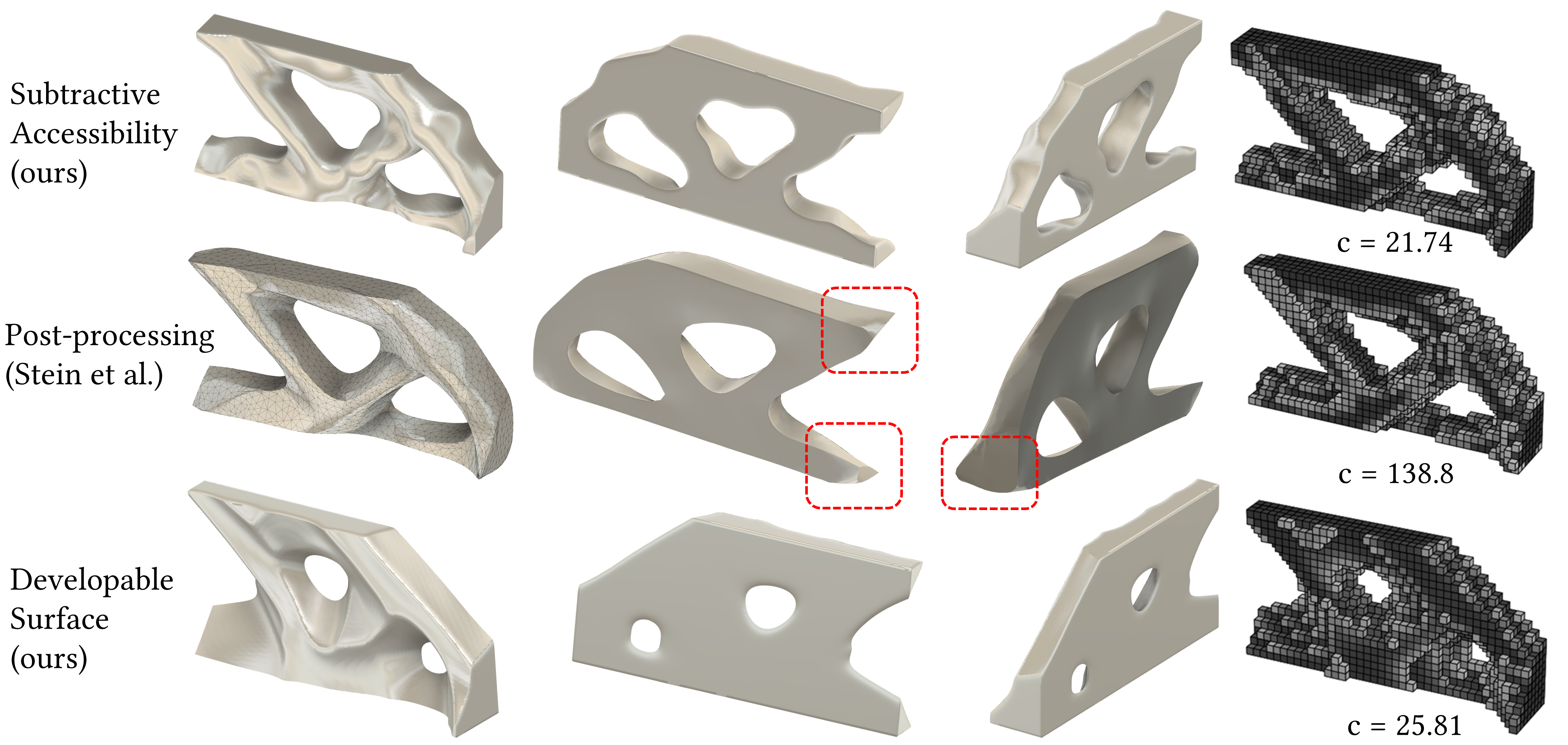}
  \caption{\ch{Comparison between Stein et al.’s post-processing method, the subtractive accessibility and developable surface result for the MBB problem. The post-processed structure exhibits developable features with distinct seamlines. However,  post-processing alters the geometry near the boundary conditions: inaccessible regions appear at the rear due to rounded corners. Regions of external forces at the front are also rounded. When all three solutions are identically voxelized for structural evaluation, these alterations negatively impact compliance.}}
  \Description{Feature size control achieved through uniform scaling}
  \label{fig:dotm}
\end{figure}

Snapshots captured during optimization (Fig. \ref{fig:ds_converge}) reveal the gradual concentration of curvature along the seamlines and the corresponding decrease in Gaussian curvature elsewhere. 

\ch{\textbf{Further lightweighting and geometry complexity.}  Fig. \ref{fig:01vf} shows the results for 400 epochs, when volume fraction is reduced to 0.1, representing a more aggressive lightweighting. While seamline formation is largely preserved, the MBB (Fig. \ref{fig:01vf}(a)) in particular exhibits a more modest delineation between the flat regions and seamlines.   We also configure a new problem (Fig. \ref{fig:01vf}(e)) with the same boundary conditions as the MBB case, with 1.25$\times$ stretching in the z direction. With the elongated z-axis, the seamline desirably follows the edges and contours of the geometry. 
}
\chh{We also evaluate the proposed method under randomized boundary conditions and varying aspect ratios with a low target volume fraction of 0.1 (Figure \ref{fig:random_bc}). The machining orientation is set to z- for the left two columns, and to both z+ and z- for the right two columns. Across these runs, the method demonstrates strong robustness.

We further compare against Autodesk Fusion 360’s commercial topology optimization solver \cite{autodesk} using identical boundary conditions and a target volume fraction of 0.3 (Figure \ref{fig:f360}). The left shows standard topology optimization, while the right shows topology optimization with a subtractive manufacturing constraint under a z- orientation. Compared to these results, we do not observe a significant difference in geometric complexity relative to our method.} \chhh{However, this comparison is limited to a single boundary-condition configuration and should not be interpreted as a comprehensive evaluation of Fusion 360. Different boundary conditions, design domains, or manufacturing settings may produce more complex geometries than those observed here.}
\ch{
\textbf{Comparison with geometric post-processing.} In Fig. \ref{fig:dotm}, we compare our results on the MBB with the surface developability post-processing approach by Stein et al. \cite{stein2018developability}. The post-processed structure by Stein et al. (Fig. \ref{fig:dotm} middle row) is obtained by applying their approach to the topology optimization with tool accessibility result shown in Figure \ref{fig:bc_compare} (first row, third column). The resulting structure (Fig. \ref{fig:dotm} middle row) exhibits characteristics similar to our developable surface (bottom row), notably bifurcations consisting primarily of three seamlines. However, since Stein et al.'s approach is geared toward transforming an existing mesh into developable pieces, machinability and displacement/force boundary condition management are understandably not the focal concern. As a result, this post-processing tends to smooth out regions critical for machining and an appropriate anchoring of the part for compliance minimization.  In this example,  post-processing smoothes out the corners at the back end of the part, resulting in regions that are not accessible with the prescribed machining orientation (Figure \ref{fig:dotm} second column). At the front of the part, the modified corners result in empty voxels at the regions of boundary conditions (third column), resulting in an expected and marked increase in compliance (fourth column). 
}

\begin{figure}[h]
  \centering
  \includegraphics[width=\linewidth]{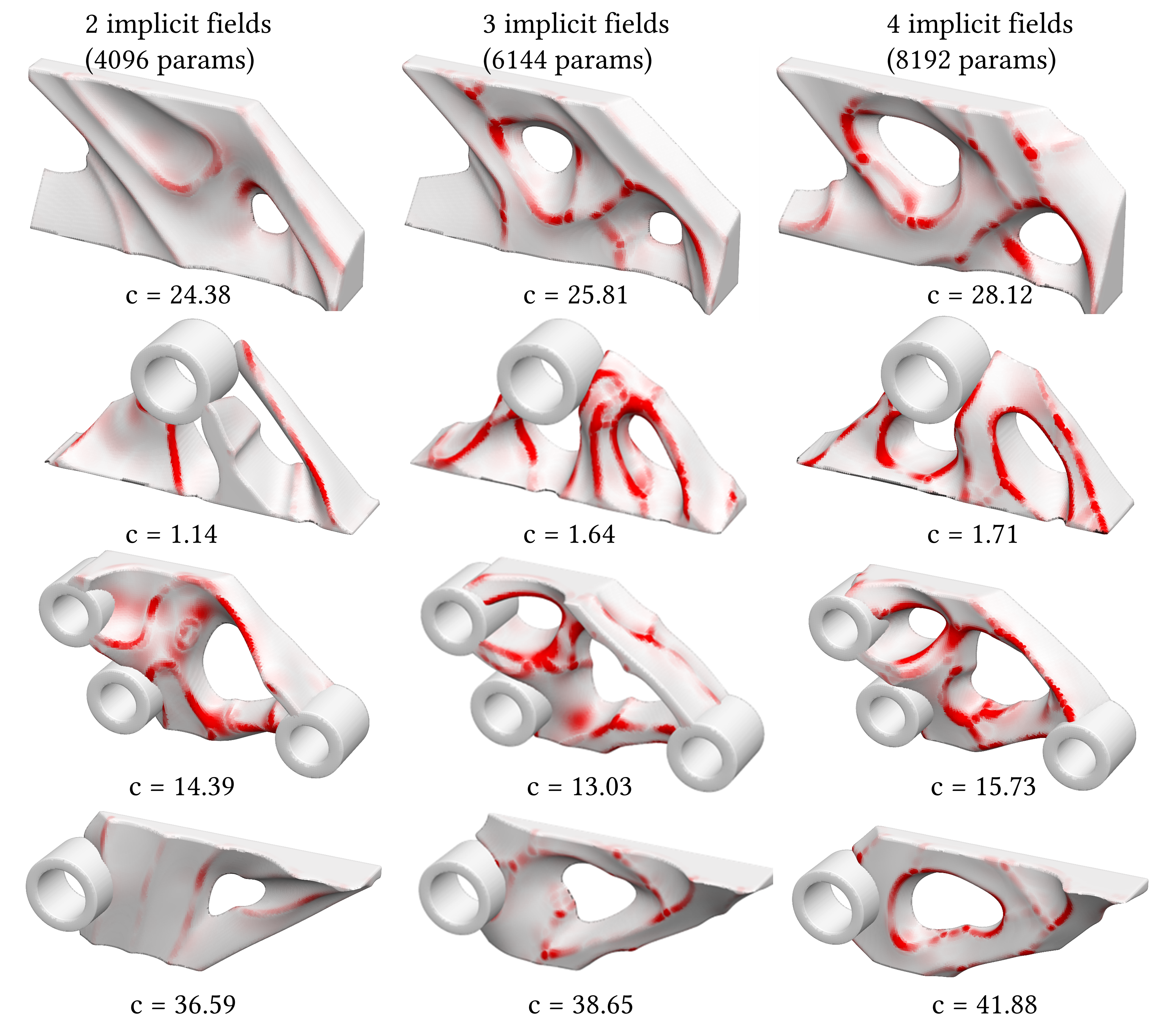}
  \caption{\ch{Results across varying number of implicit fields, together with the varying number of network parameters in each column. Optimization parameters remain identical across the runs. The only differences are the number of implicit neural fields and adjustments to seamline and intersection loss computations to accommodate the additional field. }}
  \Description{the result using three or four implicit fields}
  \label{fig:234_field}
\end{figure}

\ch{
\textbf{Number of implicit fields}. While we use three implicit fields (Fig. \ref{fig:flowchart}) throughout our examples, in this section, we showcase the results using two-, three-, and four-field implicit configurations, as shown in Figure~\ref{fig:234_field}. All optimization hyperparameters remain identical across the experiments.  With two fields, seamline trifurcation becomes impossible. This, in turn, limits how flat regions are configured and additionally increases the tendency for umbilical points, which are undesirable for flank cutting. However, compliance remains competitively low and has the least number of model parameters. The three-field configuration enables seamline trifurcations, resulting in enhanced geometric freedom, with umbilicals largely eliminated.   While it does not always achieve the best compliance, it remains competitive across the cases. Notably, the three-field is qualitatively most similar to that of  Stein et al.'s, shown previously in Figure \ref{fig:dotm}.   The four-field configuration results in slightly increased seamline complexity. In our experiments, we also observed slight issues with convergence stability in the four-field configurations, possibly due to overparameterization. Ultimately, we view the number of fields to be a hyperparameter where users must balance surface characteristics, geometric complexity,  optimization performance, and downstream use of the resulting design in light of their end goals.}


\begin{figure*}[h]
  \centering
  \includegraphics[width=\linewidth]{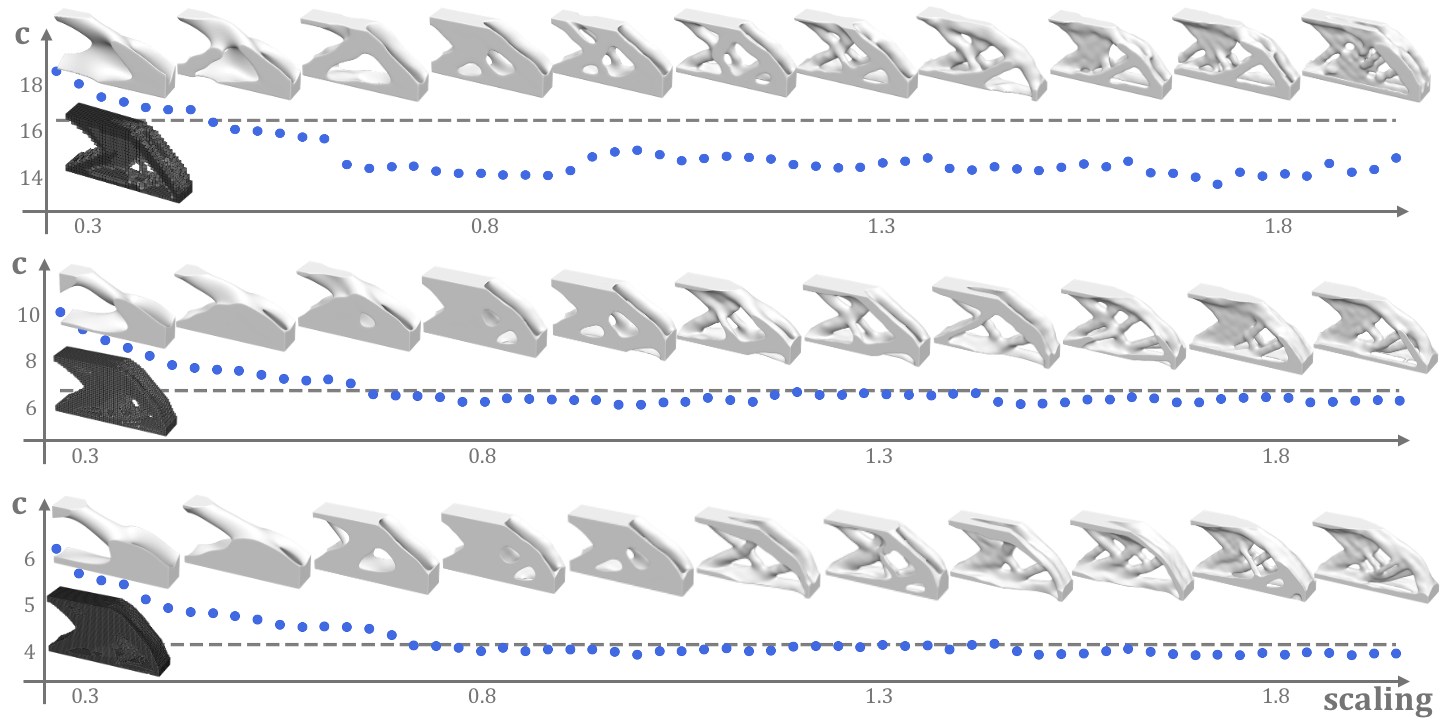}
  \caption{\chh{Feature size control in compliance minimization. We consider the same cantilever beam example at 1$\times$ (row 1), 2$\times$ (row 2), and 3$\times$ (row 3) resolutions. For each resolution, we perform 61 topology optimization runs using a \textit{single} neural network, varying the input coordinate scaling. The number of parameters remains the same for the 1$\times$ case, while the 2$\times$ and 3$\times$ cases use 8$\times$ more parameters. The resulting compliance values are shown in blue. For comparison, we also evaluate Top3D \cite{liu2014efficient} under the same boundary conditions; its compliance is shown as a dashed line, with the corresponding topology inset in the bottom left. Notably, many runs across scaling achieve lower compliance than SIMP. In contrast, increasing the resolution in SIMP does not yield a comparable increase in feature complexity under our experimental settings. As the feature scaling increases, our network produces finer geometric features.}}
  \Description{Feature size control for vanilla topology optimization}
  \label{fig:TO_freqs}
\end{figure*}

\begin{figure*}[h]
  \centering
  \includegraphics[width=\linewidth]{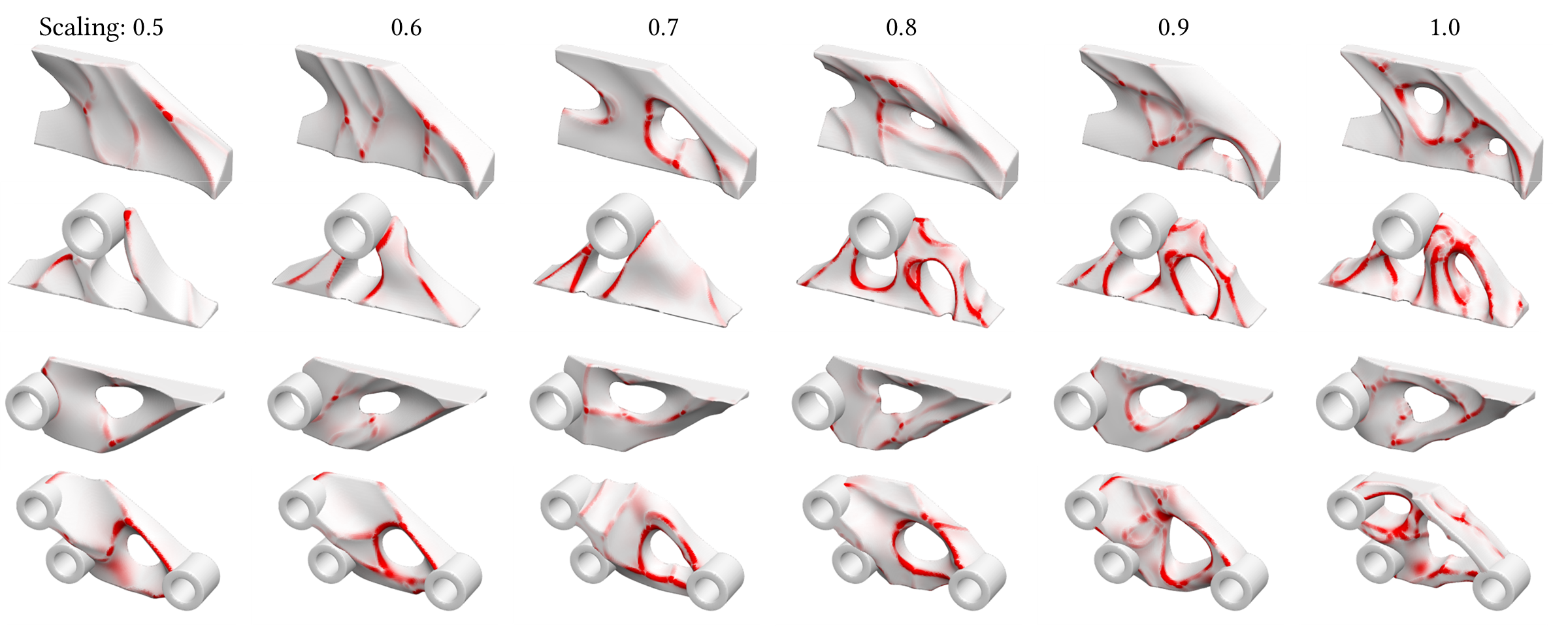}
  \caption{Feature size control achieved through uniform scaling of the input coordinates. As the scaling factor increases, the number and complexity of seamlines also increase, illustrating a trade-off between geometric intricacy and machinability. Larger scaling factors yield more intricate designs with higher seamline density, while smaller scaling factors result in simpler geometries with fewer and less complex seamlines, enhancing manufacturability.}
  \Description{Feature size control achieved through uniform scaling}
  \label{fig:gls_control}
\end{figure*}

\textbf{Feature scale control}. The developable surface results highlight the complex orientation and distribution of seamlines (Figure \ref{fig:bc_compare}). Previous studies on length scale control via implicit neural representations \cite{Chandrasekhar2021Fourier} have demonstrated that input coordinate scaling can be used to control the level of detail in the final topology.

\chh{To demonstrate the range of feature sizes achievable with our network, we consider compliance minimization on a cantilever beam across multiple resolutions (Fig. \ref{fig:TO_freqs}). For each resolution, we perform 61 topology optimization runs using a \textit{single} neural network, varying the input coordinate scaling. The parameter count is kept fixed for the 1$\times$ case and increased by 8$\times$ for the 2$\times$ and 3$\times$ resolutions.

For comparison, we evaluate Top3D \cite{liu2014efficient} under identical boundary conditions, with its compliance shown as a dashed line. Across resolutions, many runs achieve lower compliance than SIMP. Moreover, increasing the coordinate scaling consistently produces finer feature structures, whereas increasing the resolution in SIMP does not yield a comparable increase in feature complexity. These results highlight the ability of the proposed representation to control feature size through coordinate scaling while maintaining competitive compliance performance.}
\begin{figure}[h]
  \centering
  \includegraphics[width=\linewidth]{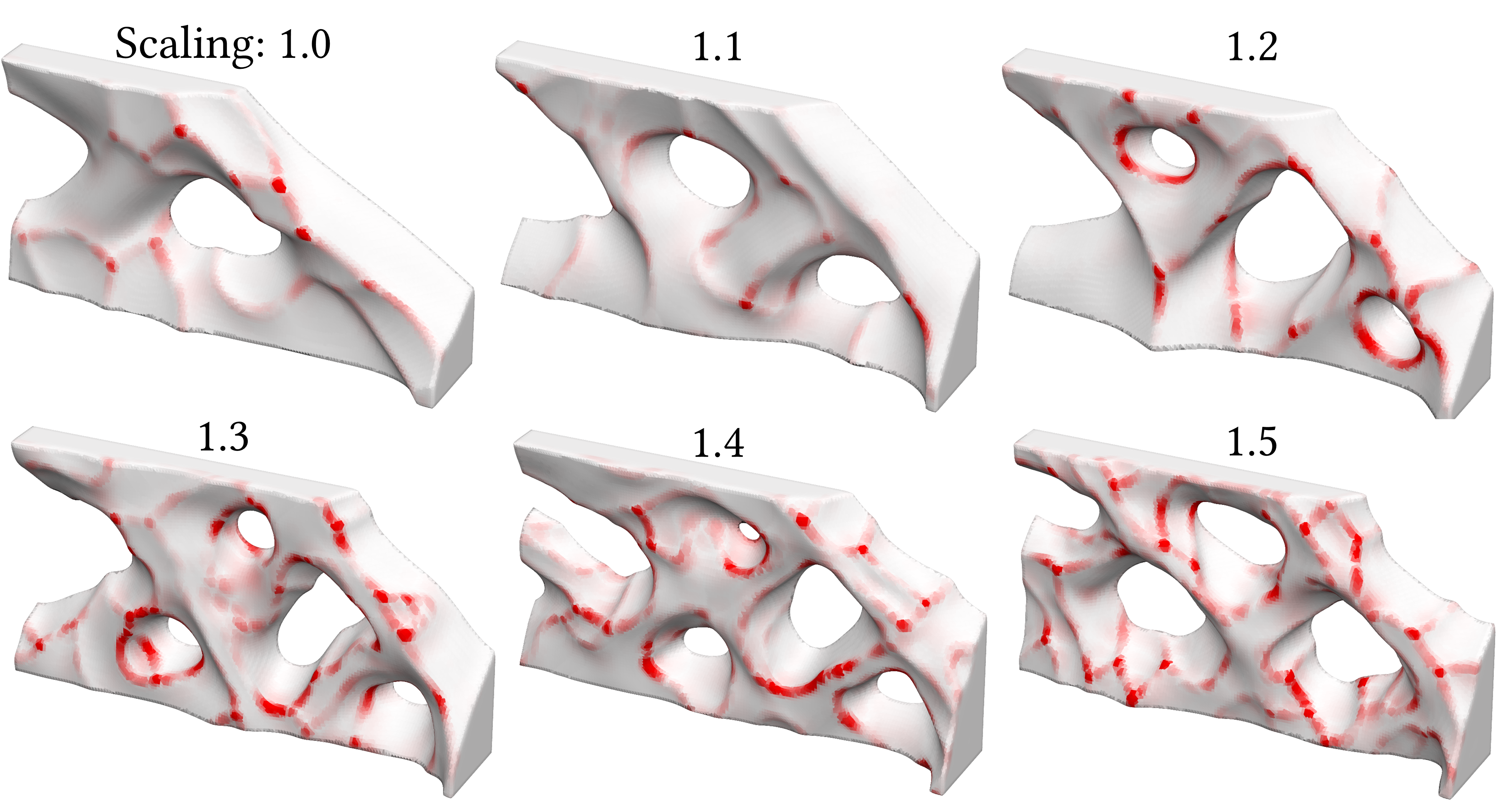}
  \caption{\ch{The finite-element and subtractive-accessibility resolutions are doubled for the MBB problem. At this higher resolution, feature scaling continues to introduce finer geometric details.}}
  \Description{Feature size control achieved through uniform scaling}
  \label{fig:feext}
\end{figure}

\ch{Next, we analyze the effect of varying input coordinate scaling from 0.5 to 1.0 across the four test cases, which now include the developability requirements. Figure \ref{fig:gls_control} shows that reduced scaling results in fewer, more continuous seamlines, while larger scales produce finer geometric features. As such, coordinate scaling enables feature size control and the exploration of a broader design space. This capability also simplifies solid modeling for practical applications, reducing complexity in CAD workflows.
}

\ch{
From the compliance minimization example (Figure \ref{fig:TO_freqs}), we note that the level of detail cannot be increased indefinitely, as the FE grid resolution eventually becomes the bottleneck. Under the subtractive accessibility constraint, Figure \ref{fig:feext} shows the additional level of detail that can be unlocked by doubling the FE resolution, at the expense of increased FE solve times. While the ability to increase geometric complexity through feature scaling and FE resolution is desirable, we note that for subsequent CAM simulations and physical machining, such intricate details can become unnecessarily prohibitive.}

\begin{figure*}[h]
  \centering
  \includegraphics[width=1.0\linewidth]{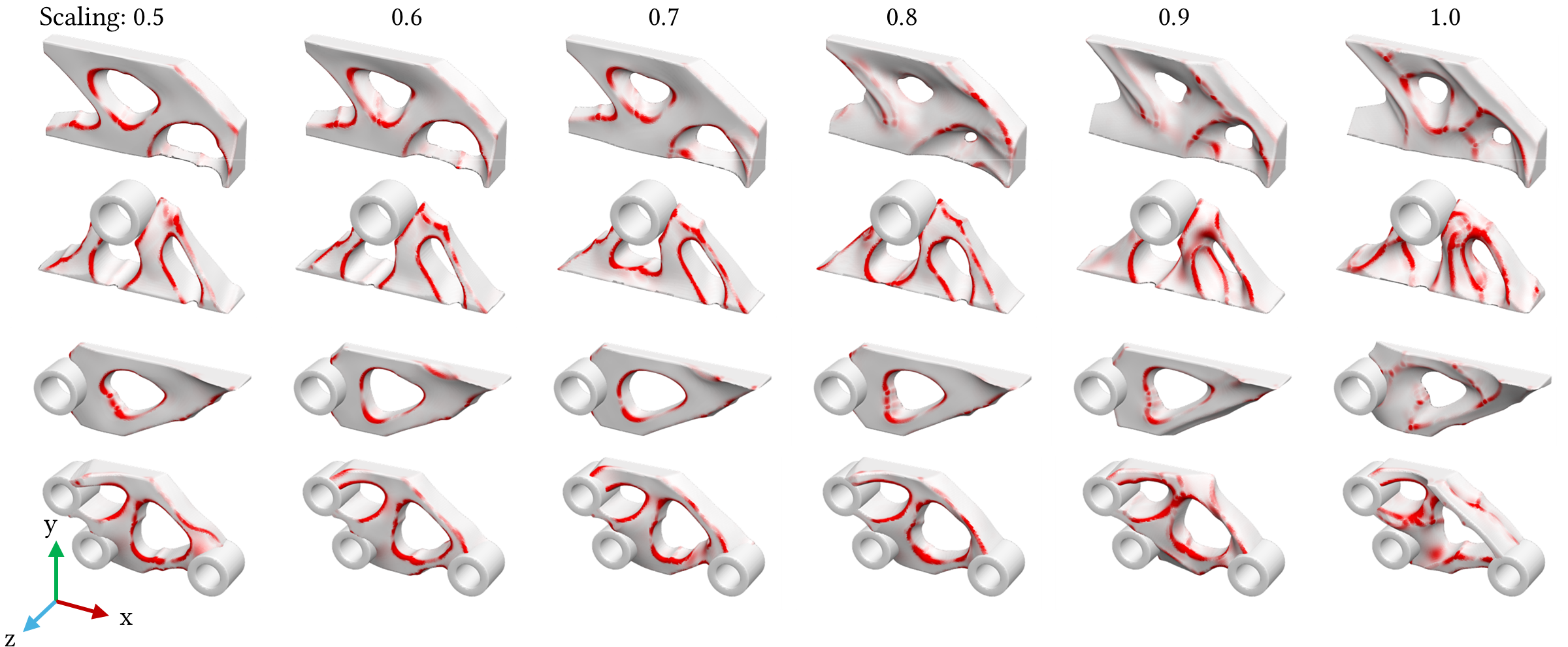}
  \caption{Variation of z-axis coordinate scaling applied to two implicit neural fields. From right to left, the z-axis scaling factor is progressively reduced. This targeted reduction results in the redistribution of seamlines, aligning them more closely with the xy-plane. The decrease in geometric complexity facilitates simplified modeling and enables more efficient configuration of CAM operations, thereby enhancing manufacturability.}
  \Description{Variation of z-axis coordinate scaling}
  \label{fig:zls_control}
\end{figure*}

We further explore feature size control by leveraging multiple neural networks with independent scaling of z-axis coordinates. As illustrated in Figure \ref{fig:zls_control}, reducing z-axis scaling aligns seamlines closer to the xy-plane, further simplifying seamline geometry compared to uniform scaling. This is achieved by scaling the z-axis coordinates of two of the neural networks while keeping the third network's z-axis coordinates unaltered. This approach facilitates solid modeling in CAD software, which often relies on construction planes for defining features.

\begin{figure}[h]
  \centering
  \includegraphics[width=1.0\linewidth]{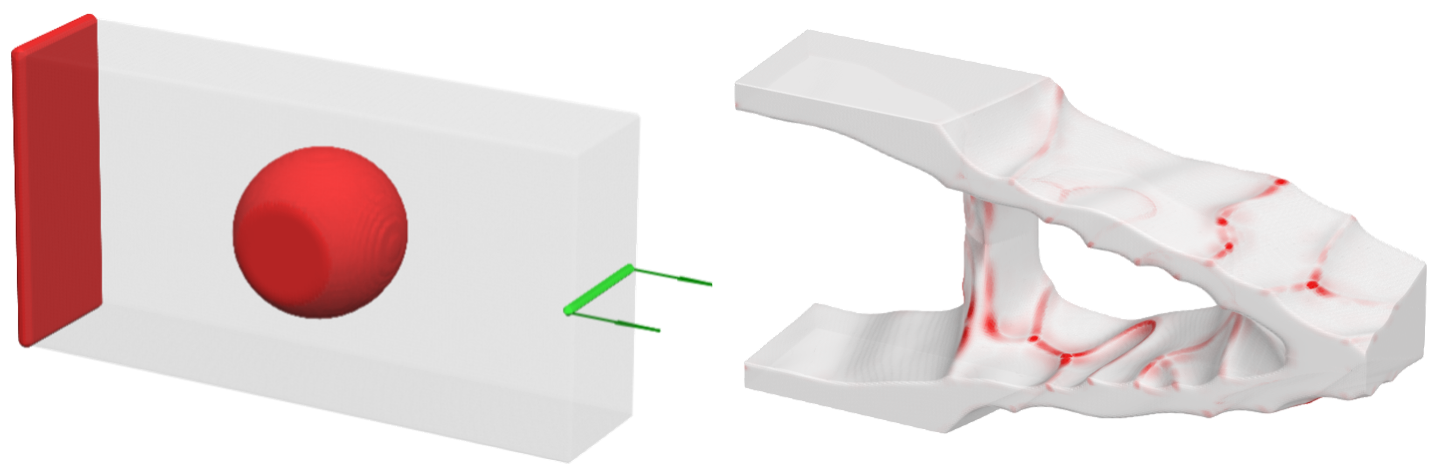}
    \caption{\chhh{Topology optimization with a spherical passive region. The left face is fixed and a distributed tensile load is applied on the right boundary, while material placement is prohibited within the central sphere. Under the $-z$ machining orientation, the optimized topology redirects the load path around the passive region while maintaining broad regions of low absolute Gaussian curvature. The geometry is mirrored about the horizontal midplane. }}
  \Description{At left, a translucent rectangular design domain contains a central spherical passive region, with a fixed red face and a green tensile load on the opposite side. At right, the optimized white structure routes material around the resulting central void; red highlights indicate regions of nonzero absolute Gaussian curvature.}
  \label{fig:no_design}
\end{figure}

\chhh{

\textbf{Passive regions.} Passive regions are prescribed subdomains in which material placement is prohibited. More specifically, we consider passive-void regions. In conventional SIMP-based topology optimization, these regions can be enforced by setting the densities of the corresponding finite elements to zero before finite element analysis \cite{liu2014efficient}. Let $\mathcal{P}$ denote the set of passive elements. The density supplied to the FE solver is then defined as

\begin{equation}\label{eqn:passive_density}
\rho_{\mathrm{FE}}^{(i)} =
\begin{cases}
0, & i \in \mathcal{P}, \\
\rho^{(i)}, & i \notin \mathcal{P},
\end{cases}
\end{equation}

\noindent where $\rho^{(i)}$ is the density produced by the neural implicit representation and $\rho_{\mathrm{FE}}^{(i)}$ is the corresponding density used for finite element analysis.

In neural implicit topology optimization, however, masking only the densities supplied to the FE solver does not constrain the underlying neural field. Consequently, $\rho^{(i)}$ may remain nonzero within $\mathcal{P}$, and material may reappear when the final implicit geometry is extracted. We therefore introduce an $L_1$ penalty on the neural-field output within the passive region:

\begin{equation}\label{eqn:passive_loss}
\mathcal{L}_{p} =
\frac{1}{|\mathcal{P}|}
\sum_{i \in \mathcal{P}}
\left|\rho^{(i)}\right|.
\end{equation}

\noindent This term is added to the combined objective with a scalar weight. Since $0 \leq \rho^{(i)} \leq 1$, minimizing $\mathcal{L}_{p}$ drives the underlying neural field toward zero throughout the passive region.

We evaluate this formulation using a rectangular design domain whose left face is fully constrained and whose right boundary is subjected to a distributed tensile load. A spherical passive region is placed at the center of the domain, obstructing the otherwise direct structure between the applied load and the fixed boundary. This configuration makes the enforcement of the passive region  apparent. Furthermore, a spherical surface has strictly positive Gaussian curvature. Simply subtracting a sphere from an optimized geometry would therefore introduce a nondevelopable surface rather than allow the optimizer to form low-curvature surfaces around it. This example evaluates whether the proposed method can simultaneously respect the prescribed passive region and produce a surrounding topology with low-curvature surfaces. We retain the same $-z$ machining orientation used in the preceding examples. As shown in Figure \ref{fig:no_design}, the optimized topology redirects the structure around the spherical region while maintaining broad regions of low Gaussian curvature.
}
\begin{figure*}[h]
  \centering
  \includegraphics[width=\linewidth]{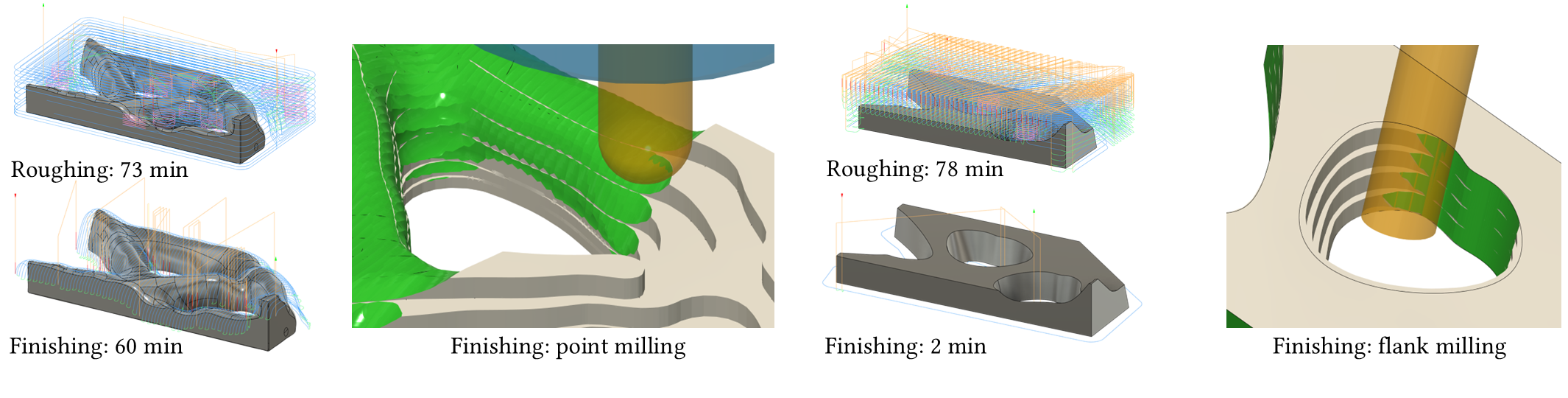}
  \caption{CAM toolpath visualization for the MBB example. Left: CAM simulation for the tool accessibility case, where the complex surface necessitates the use of a parallel toolpath, a type of point milling. This approach requires multiple traverses to approximate the surface geometry. Right: Optimized for developable surfaces. \ch{The cutting direction for flank milling is computed by the CAM software to be tangent to the surface.} This design enables the use of flank milling as the finishing toolpath, allowing the entire surface to be finished with significantly fewer passes, thereby improving machining efficiency.}
  \Description{CAM toolpath visualization for the MBB example.}
  \label{fig:sub_ds_mbb_cam}
\end{figure*}

\subsection{CAM Analysis and Machining Time Comparison}

To further evaluate manufacturability, we performed CAM analysis using Autodesk Fusion 360 \cite{autodesk}. We set the scaling factor between FE element size and the physical part dimensions to be $0.1 \text{ in/element}$, hence making, for instance, the MBB beam's physical dimensions $4 \text{ in} \times 2 \text{ in} \times 0.8 \text{ in}$ (refer to Table \ref{tab:bc_compare}). For topology optimization with tool accessibility, we employed the marching cubes algorithm to generate a mesh at the 0.5 level set, which was subsequently imported into Fusion 360 and converted into a solid using the \textit{convert mesh} function. For the optimization of developable surfaces, a test case with a reduced z-axis feature size was manually modeled in Fusion 360 for similarity and ease of manufacturing.

Both examples began with rough machining, a process focused on high material removal rates without prioritizing surface finish. This process involves high material removal rates and higher cutting depths, speeds, and feeds, aiming to shape the workpiece roughly to its near-final form in preparation for finishing operations. The \texttt{pocket cleaning} strategy in Fusion 360 was used for rough machining, generating offset contours to rapidly remove bulk material. Finishing operations employed different strategies based on geometry: for tool accessibility cases with complex curved surfaces, a ball-nose endmill with the \texttt{Parallel} strategy was utilized, while for developable surfaces, flank milling (or Swarf machining) with a flat-head endmill was employed. Figure \ref{fig:sub_ds_mbb_cam} provides CAM toolpath visualizations for the MBB beam case study.

For consistency, the cutting feed rate and spindle speed were maintained across rough machining operations. Similar mill speed parameters were targeted for finish machining to ensure a fair comparison. Detailed machining parameters are summarized in Table \ref{tab:CAM_config}.

\begin{table}[h!]
\centering

\begin{tabular}{l|c|c|c}
\textbf{Parameter} & \textbf{Roughing} & \textbf{Parallel} & \textbf{Flank} \\
\hline
\textbf{Tool (3/8 Dia)} & Flat Head & Ball Nose & Flat Head \\
\textbf{Spindle Speed (RPM)} & 3000 & 3000 & 3000 \\
\textbf{Feedrate (in/min)} & 10 & 10 & 10 \\
\textbf{Load/Stepover (in)} & 0.1 & 0.02 & 0.1 \\
\textbf{Stepdown (in)} & 0.1 & 0.1 & 0.1 \\
\end{tabular}
\caption{Configuration of machining strategies for roughing, parallel, and flank operations. The selected parameters are intentionally conservative, reflecting the limitations of the relatively small CNC mill utilized in this study. To ensure consistency and comparability, the same machining parameters are applied when fabricating physical examples.}
\label{tab:CAM_config}
\end{table}

CAM toolpaths were generated based on specified parameters using Autodesk Fusion 360. Machining time estimates, derived from these toolpaths, are summarized in Table \ref{tab:merged_cam} for both optimizations for tool accessibility and developable surfaces across the four test cases. Roughing times were comparable between the two approaches, attributable to the similar volume fractions achieved during topology optimization. This results in largely similar bulk material removal rates across test cases. However, significant differences were observed during finishing operations. Topology optimization incorporating tool accessibility required longer finishing times, primarily due to the increased travel distance of the ball nose endmill necessary to achieve acceptable surface finishes. In contrast, the developable surface approach demonstrates substantial reductions in finishing time. This is because the cutter primarily moves along the tangential direction of the developable surface, requiring fewer passes and shorter travel distances.

\begin{table*}[h!]
\centering

\begin{tabular}{l|ccc|ccc|ccc}
\textbf{Problem Type} & \multicolumn{3}{c|}{\textbf{Subtractive accessibility$\Lc$ $\Lm$}} & \multicolumn{3}{c|}{\textbf{Developable surface$\Lc$ $\Lm$ $\Lg$}} & \multicolumn{3}{c}{\textbf{Cutter-surface align$\Lc$ $\Lm$ $\Lg$ $\La$}} \\
 & \textit{Roughing} & \textit{Finishing} & \textit{Total} & \textit{Roughing} & \textit{Finishing} & \textit{Total} & \textit{Roughing} & \textit{Finishing} & \textit{Total} \\
\hline
\textbf{MBB} & 73 & 60 & 133 & 78 & 2 & 80 & 89 & 1 & 90 \\
\textbf{Mount} & 91 & 44 & 135 & 82 & 3 & 85 & 85 & 3 & 88 \\
\textbf{Pedal} & 111 & 63 & 174 & 99 & 2 & 101 & 112 & 1 & 113 \\
\textbf{Link} & 76 & 51 & 127 & 65 & 3 & 68 & 74 & 4 & 78 \\
\end{tabular}
\caption{Machining times for roughing, finishing, and all operations combined for the subtractive-accessibility, developable-surface, and cutter-surface-alignment approaches. All times are in minutes.}
\label{tab:merged_cam}
\end{table*}

\subsection{3-axis milling tailored optimization with cutter-surface alignment}

\begin{figure}[h]
  \centering
  \includegraphics[width=\linewidth]{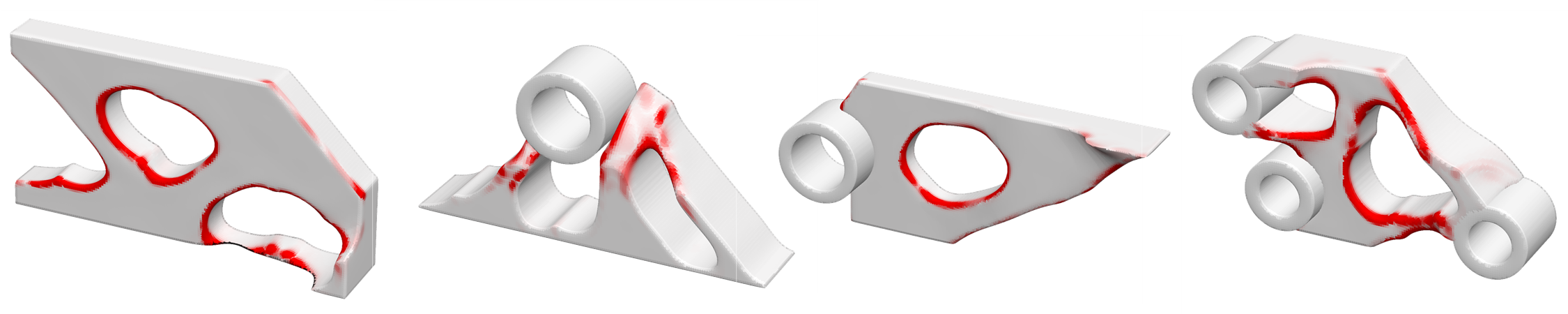}
  \caption{Four test cases with cutter-surface alignment loss $\mathcal{L}_a$ enabled.  }
  \Description{Four test cases with cutter-surface alignment loss enabled.}
  \label{fig:sub_a35_compare}
\end{figure}

\begin{figure}[h]
  \centering
  \includegraphics[width=\linewidth]{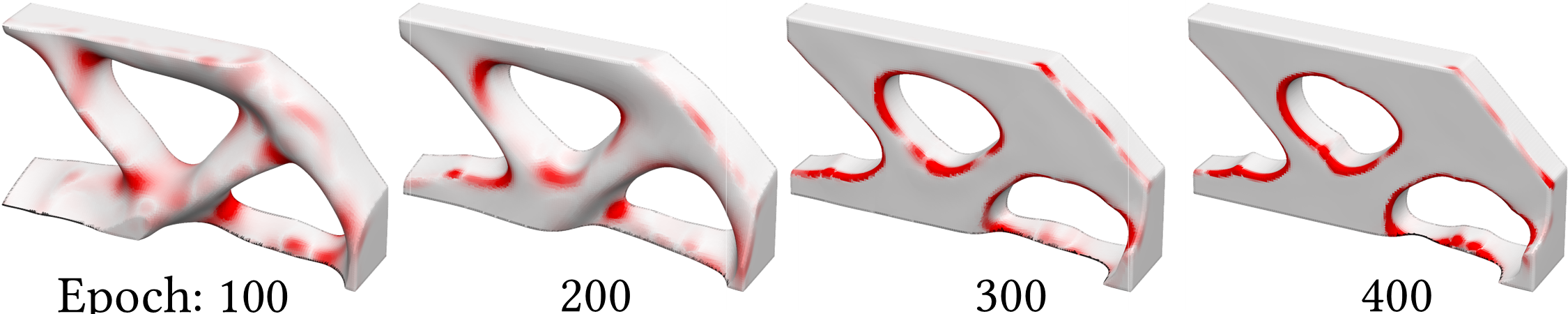}
  \caption{Convergence history for the mbb beam example with alignment loss enabled.  }
  \Description{Convergence history for the mbb beam example with alignment loss enabled}
  \label{fig:a3_converge}
\end{figure}

Despite the machining efficiencies of developable surfaces, flank milling these geometries necessitates the use of 5-axis machining centers, which are less prevalent than 3-axis configurations \cite{dmgmori_5axis}.  As described previously, we introduce an alignment loss function ($\mathcal{L}_a$) to make shape design amenable to   3-axis milling. Observations from prior results showed that reducing z-axis scaling constrained seamlines closer to the xy-plane, aligning surfaces with the cutter's tangential and normal orientations. To validate this approach, we reduced the z-axis scaling to 0.5 and applied the alignment loss across all test cases. The resulting geometries, visualized in Figure \ref{fig:sub_a35_compare}, exhibit enhanced alignment of surfaces with the 3-axis milling cutter compared to previous developable surface results (Figure \ref{fig:zls_control}).

\begin{figure}[h]
  \centering
  \includegraphics[width=\linewidth]{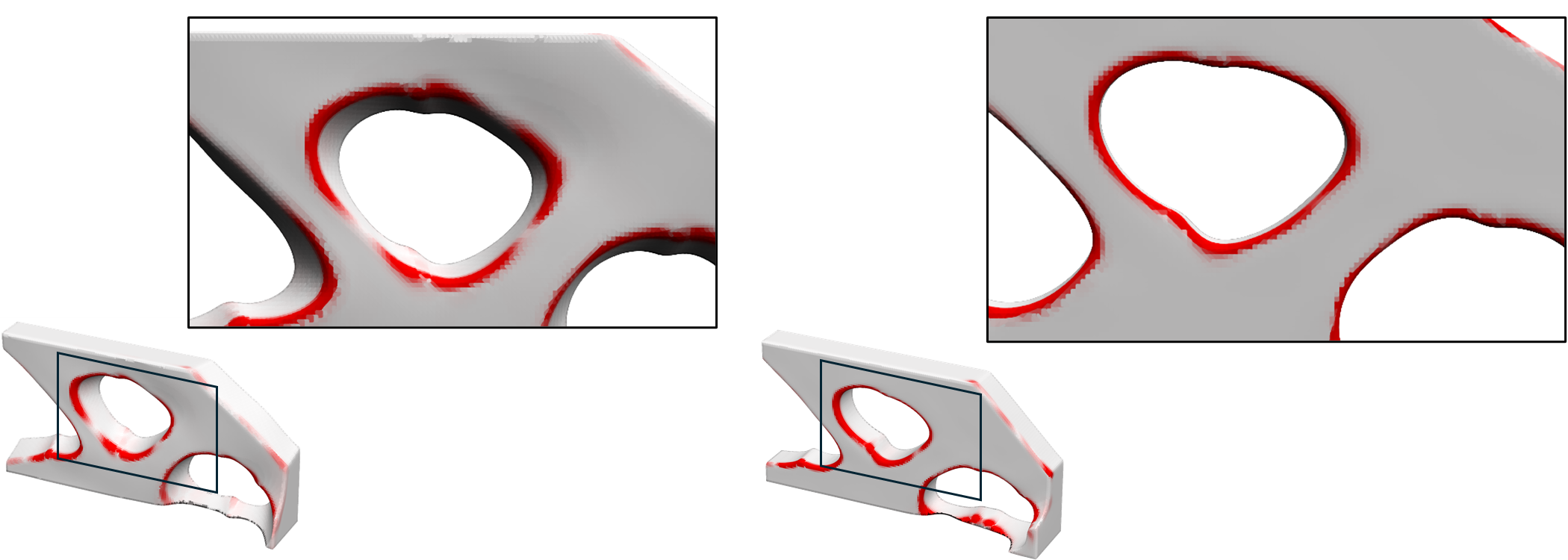}
  \caption{Close-up comparison between the developable surface (left) and the 3-axis milling-oriented (right) results. While the overall geometric similarities are evident, the 3-axis milling-oriented design exhibits improved surface alignment with the cutter, facilitating enhanced compatibility with 3-axis machining.  }
  \Description{Close-up comparison between the developable surface and the 3-axis milling-oriented results.}
  \label{fig:a35_zoom}
\end{figure}

Convergence during optimization is shown in Figure \ref{fig:a3_converge}, illustrating how the surface gradually aligns with the milling tool. A close-up comparison in Figure \ref{fig:a35_zoom} highlights the improved cutter-surface alignment achieved through this approach.

The 3-axis-optimized geometries were modeled and subjected to CAM analysis. The alignment of surfaces with the xy-plane facilitated simplified solid modeling, as the geometries could be approximated by extruding sketches on reference planes. Additionally, the roughing operation closely approximated the final geometry due to the alignment of most surfaces with the cutter's tangential and normal directions. However, rounded fillets along edges required finishing operations using a ball nose endmill, categorized as finishing operations. Consistent machining configurations, as summarized in Table \ref{tab:CAM_config}, were used across all cases. The estimated machining times are detailed in Table \ref{tab:merged_cam}.

To validate the CAM simulations, we machined two MBB beam examples—one optimized for tool accessibility and the other for 3-axis milling—using Aluminum 6061 on a Haas OM-2 machine \cite{haas_compact_mills}. Estimated machining times from Fusion 360 were compared to actual machining times, summarized in Table \ref{tab:sub_a3_actual}. Results demonstrate a very close agreement between estimated and actual times, reinforcing our trust in the CAM simulations in this presented context. 

Notably, the 3-axis-optimized example required slightly more roughing time due to the closer resemblance of the roughing geometry to the final part, which involved marginally higher material removal. For both examples, a uniform base was added to facilitate fixturing during machining and was manually removed afterward using a mill.

\begin{table*}[h!]
\centering

\begin{tabular}{l|ccc|ccc}
\textbf{Problem Type} & \multicolumn{3}{c|}{\textbf{Subtractive accessibility $\Lc$ $\Lm$}} & \multicolumn{3}{c}{\textbf{Cutter-surface align $\Lc$ $\Lm$ $\Lg$ $\La$}} \\
 & \textit{Roughing} & \textit{Finishing} & \textit{Total} & \textit{Roughing} & \textit{Finishing} & \textit{Total} \\
\hline
\textbf{mbb (CAM estimated)} & 73 & 60 & 133 & 89 & 1 & 90 \\
\textbf{mbb (actual)} & 75 & 62 & 137 & 92 & 1 & 93 \\

\end{tabular}
\caption{Comparison of estimated machining time with actual machining time on the Haas OM2 mill. The results demonstrate that the Fusion 360 time estimation closely matches the actual machining duration. During the machining process, the feed rate was reduced on a few occasions when the machine indicated a heavy spindle load or when the long ball nose endmill produced chattering, which could potentially affect the quality of the machining process.}
\label{tab:sub_a3_actual}
\end{table*}

\begin{figure}[h]
  \centering
  \includegraphics[width=\linewidth]{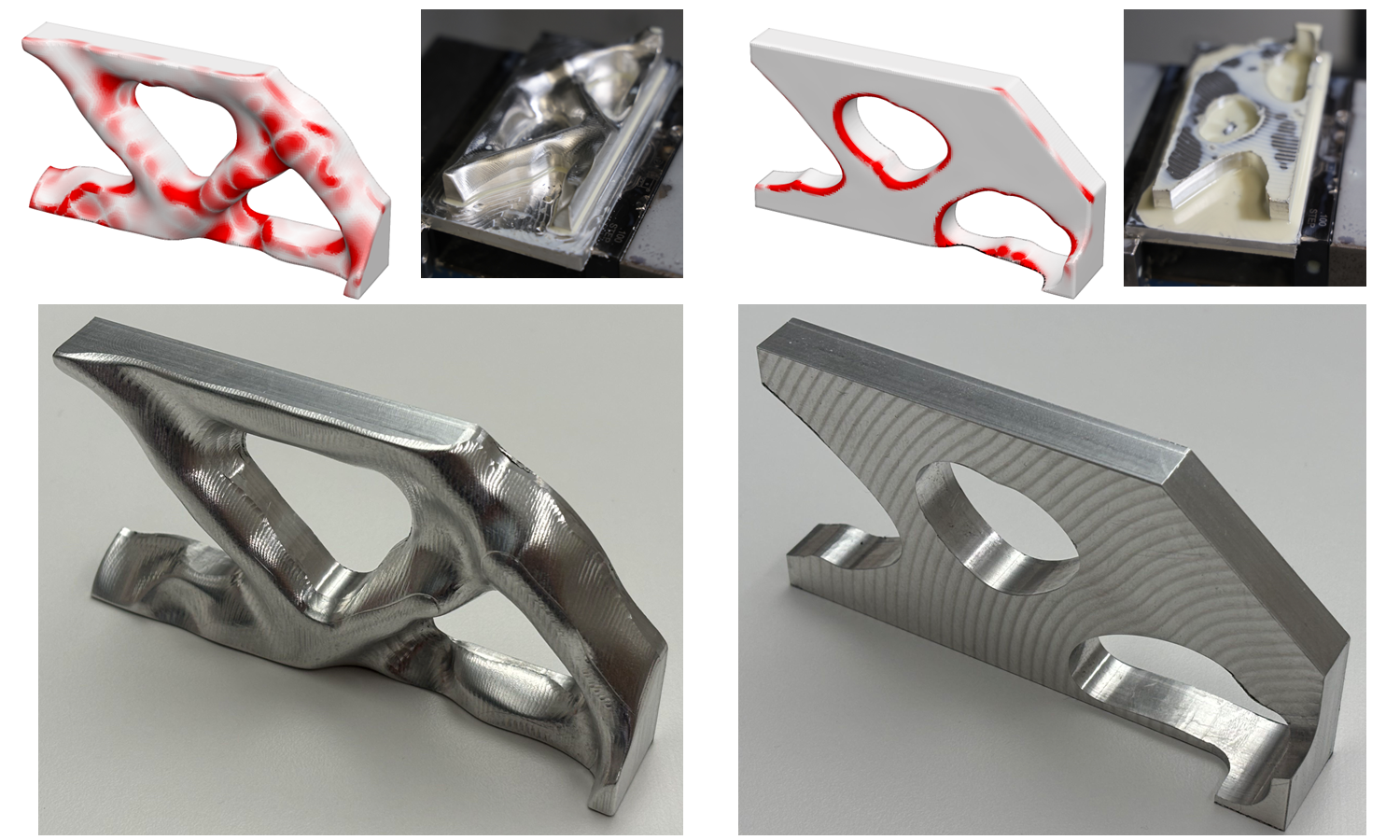}
  \caption{Physical machining of the two examples was performed on a Haas OM2 using Aluminum 6061. Left: optimized for tool accessibility. Right: optimized for cutter-surface alignment (3-axis milling).  }
  \Description{Physical machining of the two examples}
  \label{fig:sub_a3_cam_part}
\end{figure}

Figure \ref{fig:sub_a3_cam_part} shows the optimized design and the finished part. We can see an overall smoother surface finish on the 3-axis tailored design. 

\section{Limitations and Future Work}

This study presents a potential pathway for optimizing geometries for subtractive manufacturability by leveraging developable surfaces. While promising, several challenges remain in the practical machining of developable surfaces, particularly when using flank milling techniques, as highlighted in prior research \cite{harik2013flank}. The integration of implicit neural representations offers opportunities to compute additional surface properties. For example, these representations could facilitate the inclusion of tool diameter constraints alongside principal curvature constraints, ensuring the cutter's orientation remains within a predefined motion envelope. Furthermore, adopting strategies to prevent umbilical points using formulations derived for implicit surfaces \cite{che2007lines} could further enhance manufacturability.

Currently, the conversion of optimized results into manufacturable geometries depends on manual processes. Prior research has explored automated methods for converting b-rep to implicit neural representations \cite{guo2022implicit}, as well as mesh \cite{shen2025mesh2brep} and point cloud \cite{li2023surface} data into b-rep formats. Future research must build on these advancements to develop algorithms capable of faithfully converting implicitly defined solids with sharp seamlines into b-rep representations. Such algorithms must also preserve the developable nature of the surfaces, as current formulations face limitations in generating convex seams. This constraint, while beneficial in some scenarios such as avoiding the use of very small-diameter endmills, presents an area for improvement in enabling more complex geometric designs.

While compliance minimization remains a widely adopted strategy in topology optimization, more advanced formulations, such as stress-constrained mass minimization, could significantly broaden the applicability of this work. Similarly, incorporating additional physical criteria, including fatigue resistance, thermal-fluidic interactions, or optical performance, would extend the method's utility to more complex engineering applications.

Lastly, tailoring geometry representations to accommodate specific cutter geometries and optimizing motion planning for efficient manufacturing operations \cite{bartovn2021geometry} represent promising directions for future research. These efforts would ensure that the optimized designs align with practical machining constraints, further bridging the gap between computational optimization and real-world manufacturability.


\section{Conclusion}

We introduce a new method that integrates topology optimization with manufacturability constraints to address the challenges associated with the subtractive machining of complex geometries. By leveraging implicit neural fields, we gain precise control over surface developability, enabling the generation of designs characterized by localized seamlines that concentrate curvature while maintaining regions of the geometry that remain developable. These seamlines not only facilitate efficient toolpath generation for subtractive processes such as 5-axis flank milling but also enable the adaptation of designs to meet the constraints of 3-axis machining.

Our findings demonstrate that this approach leads to a reduction in machining time compared to conventional methods, without significantly compromising structural performance. By achieving an optimal balance between design flexibility and machining efficiency, this methodology bridges the gap between the complex geometries generated by topology optimization and the practical requirements of subtractive manufacturing. The tri-implicit field representation offers a versatile framework for regulating curvature distribution and aligning geometries with specific machining constraints. This advancement paves the way for the development of more efficient and manufacturable designs in both additive and subtractive manufacturing contexts.

\section*{Acknowledgement}
We would like to thank Professor Jim McCann for his valuable feedback and support throughout this work.

\bibliographystyle{ACM-Reference-Format}
\bibliography{references}

\end{document}